\documentclass[twocolumn]{aastex701}

\usepackage{amsmath}

\shorttitle{Eruptive novae in symbiotic systems-extended analysis}
\shortauthors{I. B. Vathachira et al}

\begin{document}

\title{The role of binary configuration in shaping nova evolution via wind accretion in symbiotic systems}

\author[0009-0003-3561-5961]{Irin Babu Vathachira}
\affiliation{Department of Physics, Ariel University, Ariel, POB 3, 4070000, Israel}
\email[show]{irinbabu00@gmail.com}  

\author[0000-0002-0023-0485]{Yael Hillman}
\affiliation{Department of Physics, Azrieli College of Engineering Jerusalem, Israel}
\email[show]{yaelhl@jce.ac.il}

\author[0000-0002-7840-0181]{Amit Kashi}
\affiliation{Department of Physics, Ariel University, Ariel, POB 3, 4070000, Israel}
\affiliation{Astrophysics, Geophysics and Space Science (AGASS) Center, Ariel University, Ariel, 4070000, Israel}
\email[show]{kashi@ariel.ac.il}

\begin{abstract}

We investigate the impact of the Bondi-Hoyle-Lyttleton (BHL) accretion mechanism on the evolution of nova eruptions in symbiotic systems, by systematically varying three key input parameters: the initial donor (asymptotic giant branch, AGB) mass, the initial white dwarf (WD) mass, and the initial binary separation ($a$). We explore models with AGB masses in the range $1.5-3.5~\rm{M_\odot}$; WD masses in the range $0.7-1.25~\rm{M_\odot}$ and separations in the range $1-8~k\rm{R_\odot}$.
We find all our models to show a significant, long-term orbital increase. This trend is primarily driven by the fact that $\sim99\%$ of the AGB’s mass is lost from the system, either directly via wind --- that is never accreted onto the WD --- or accreted onto the WD and then ejected during nova eruptions. This results in the effect of the mass loss (or transfer) on the orbit to dominate over the effect of the angular momentum loss sinks that could shrink the orbit, leading a consistent orbit widening. Consequently, all of our WD masses gradually decrease.
A more massive WD achieves a higher mass transfer efficiency and accretion rate, meaning a slightly better mass retention efficiency per nova. However, since a higher accretion rate causes more frequent eruptions, the total WD mass loss over the AGB lifetime, is more substantial. 
We conclude that symbiotic systems transferring mass via the BHL mechanism are unlikely to be type Ia supernovae progenitors.

\end{abstract}


\section{Introduction}
Binary systems hosting white dwarfs (WDs) stand as potential candidates for novae. The hydrogen-rich material accreted from the binary donor accumulates as a shell enveloping the WD. As additional matter is accreted, the accumulated material undergoes compression, resulting in elevated pressure and temperature at the base of the envelope (e.g., \citealt{1976MNRAS.176...53G,1978ARA&A..16..171G,2011arXiv1111.4941B,2016PASP..128e1001S,2021ARA&A..59..391C}). Once a critical amount of mass is accreted, the conditions initiate nuclear burning, which will rapidly develop and trigger a thermonuclear runaway (TNR), causing the accreted envelope to be expelled from the WD's surface \cite[] {1972ApJ...176..169S}.
The ejection of the accreted material due to a TNR is commonly labeled as a nova eruption (e.g., \citealt{1971MNRAS.152..307S,1972ApJ...176..169S,1978A&A....62..339P,1978ApJ...226..186S,1986ApJ...310..222P,2016PASP..128e1001S,2022ApJ...938...31S}), although novae can also occur exceptionally rarely without the expulsion of matter (e.g., \citealt {1995ApJ...445..789P,2005ApJ...623..398Y,2007ApJ...660.1444S,2012BASI...40..419S,2013ApJ...777..136W,2019ApJ...879L...5H,2024MNRAS.527.4806V}). Nevertheless, both types of eruptions exhibit a surge in brightness, reaching levels of several tens of thousands of times the Sun's brightness, followed by a gradual decline over time  (e.g., \citealt{1964gano.book.....P,1978ARA&A..16..171G}). The mechanism of accretion-ejection persists as long as the donor still has mass to supply, i.e., until it is eroded \cite[]{2021MNRAS.501..201H,2021MNRAS.505.3260H}, or until the WD gains enough mass to reach the Chandrasekhar limit and undergo a type Ia supernova explosion (SNIa) (e.g., \citealt{1982ApJ...253..798N,1997ApJ...475..740N,2013ApJ...767...57F,2016ApJ...819..168H,2017hsn..book.1275N}). 

The most common way of classifying novae is to categorize them as classical novae (CNe) and recurrent novae (RNe). This classification hinges solely on the time between two successive eruptions (i.e., the recurrence time,  $\mathrm{t_{rec}}$). CNe typically exhibit longer recurrence times, by definition surpassing a century, while RNe are identified by having at least two observed eruptions or a $\mathrm{t_{rec}}$ of less than a century \cite []{2014ApJ...788..164P}. Symbiotic novae constitute binary systems that comprise an accretor WD and an evolved donor --- A red giant (RG) or asymptotic giant branch star (AGB), with the WD typically accreting matter from the wind of the evolved donor (e.g., \citealt{1983ApJ...273..280K,2021MNRAS.501..201H}). In principle, a symbiotic system (SyS) can host both CNe and RNe. The two parameters that have been established as the key influencers on the time between eruptions is the WD mass ($M_{\rm WD}$) and the average accretion rate ($\dot{M}_{\rm acc,avg}$). A higher average accretion rate will accrete the triggering mass in a shorter time period, and a more massive WD will require a lower triggering mass due to the high surface gravity, thus $t_{\rm rec}$ decreases with increasing $M_{\rm WD}$ or increasing $\dot{M}_{\rm acc,avg}$ (e.g., \citealt{1995Ap&SS.230...75S,2008ASPC..401..131T,2012BaltA..21...76S}).
For this reason, CNe typically host less massive WDs than RNe \cite[] {2005ApJ...623..398Y}. This has been demonstrated for a pool of 82 Galactic CNe for which the average WD mass is $\sim1.13 \rm M_{\odot}$, and after compensating for eruption frequency the average WD mass is $\sim1.06 \rm M_{\odot}$, whereas the average WD mass in the 10 known Galactic RNe, is $\sim1.31 \rm M_{\odot}$ \cite[]{2018ApJ...860..110S}.

When considering giant donors, wind and mass loss becomes a crucial factors in their life cycle. Among them, AGBs emerge as an interesting class of donors due to their powerful stellar winds, resulting in mass loss rates roughly in the range 
$\mathrm{10^{-7}-10^{-5}M_\odot yr^{-1}}$ \cite []{2009ASPC..414....3H}. These AGB stars exhibit luminosity variations caused by thermal pulses, which result from alternating helium and hydrogen-shell burning, inducing alterations in radius and temperature that impact the overall luminosity of the star (e.g., \citealt{1983A&A...127...73B,1993ApJ...413..641V,dorfi1998agb}). Thermal pulses in AGB stars generate shock waves that lift gas outward. As the gas rises above the condensation radius, it cools below the dust condensation temperature, allowing dust grains to form. Radiation pressure on these dust grains then accelerates both the dust and surrounding gas, and if they reach velocities exceeding the star’s escape velocity, they are driven away from the stellar surface as a stellar wind. (e.g., \citealt{dorfi1998agb,2009ASPC..414....3H}). The stellar mass loss rate varies due to fluctuations in both radius and density during thermal pulses and is assumed to flow out of the star isotropically.

In SySs with long orbital periods, i.e., a relatively wide binary separation, the stellar components will be detached (e.g., \citealt{1999A&AS..137..473M,2019CoSka..49..189S}).
Symbiotic binaries with S-type donors (RGBs) typically have an orbital period of a few years (e.g., \citealt{2000A&AS..146..407B,2013AcA....63..405G}), whereas for D-type (Mira) donors (AGBs), the orbital period can extend much longer (e.g., \citealt{1975MNRAS.171..171W,2013ApJ...770...28H,2025A&A...695A..61M}). In these systems, the possibility of Roche lobe overflow (RLOF) or gravitational focusing of wind is ruled out (e.g., \citealt{2013A&A...552A..26A,2018MNRAS.473..747C,2025ApJ...980..224V}), thus, Bondi-Hoyle-Lyttleton (BHL) accretion emerges as the favorable form of mass transfer to the WD. The BHL mechanism demonstrates how a point mass embedded in a cloud of gas accretes matter and is commonly used to describe wind accretion  \cite[]{1939PCPS...35..405H,1944MNRAS.104..273B,1952MNRAS.112..195B}. In this scenario, the isotropically escaping wind from the donor forms a cloud of matter, and the WD acts as a point object within this cloud, accreting a portion of wind, while the rest is lost from the system. The material that escapes from the system carries with it a specific angular momentum, influencing the binary separation of the system.

In this study, we investigate the nature and characteristics of multiple consecutive nova eruptions as the result of accretion from the wind of AGB donors via the BHL mechanism. This work is a direct continuation of previous studies \cite[] {2021MNRAS.501..201H, 2024MNRAS.527.4806V}, which investigated eruptions on different WDs and binary separations while assuming a single initial AGB donor model with a mass of $1.0\mathrm{M_\odot}$. In the present study, we have considered various initial mass AGB models to investigate both the effects of the WD mass ($M_{\rm WD}$) and the donor AGB mass ($M_{\rm AGB}$), as well as their binary separation ($a$), on the behavior of novae and the system's evolution. 

In the next section we explain our computation methods and specify our models.  Our results are presented in \S \ref{sec:results}, followed by a discussion in \S \ref{sec:Discussion} and our conclusions in \S \ref{sec:Conclusions}.

\begin{table*}
\hspace{-0.5cm}
\begin{tabular}{c c c c c c c c c}
\hline
          &  &  input &   &   & &  output& &   \\
\hline
         Model & $M_{\rm AGB,ini}$   &  $M_{\rm WD,ini}$ & $ a_{\rm ini}$ [$10^3$]  & $P_{\rm orb,ini}$ & Number of &  $\Delta M_{\rm WD}$ & $ \Delta a $ &  $\Delta P_{\rm orb}$\\

         $\#$& [$\rm{M_{\odot}}$]  &  [$\rm{M_{\odot}}$] & [$\rm{R_{\odot}}$] &  [yrs]&  cycles &[\%] & [\%] & [\%]  \\
\hline
\\
\multicolumn{9}{c}{\textbf{Varying AGB mass}} 
\\
\hline
        { }&{}&{}&{}&{}&{ }&{}&{}&{}\\
        1& 1.42& 1.0  & 8.0   &145.7 & 40 &  -0.013 & +233 &  +651 \\
        2& 2.49& 1.0  & 8.0   &121.3 & 74 &  -0.022 & +460 &  +1841\\
        3& 3.49& 1.0  & 8.0   &107.0 & 99 &  -0.005 & +766 &  +4029\\
        {}&{}&{}&{}&{}&{}&{}&{}&{}\\
      
\hline
\\
\multicolumn{9}{c}{\textbf{Varying Separation}} 
\\
\hline        
          
          {}&{}&{}&{}&{}&{}&{}&{}&{}\\
	  $4^{*}$ &1.42 &1.0  & 1.0  & 6.4  & 3093 & +1.140& +135 &  +343\\
        5 &1.42 &1.0  & 4.0  & 51.5 & 133  & -0.018& +219 &  +605\\
        6 &1.42 &1.0  & 8.0  & 145.7& 40   & -0.013& +233 &  +651\\
        {}&{}&{}&{}&{}&{}&{}&{}&{}\\
\hline
\\
\multicolumn{9}{c}{\textbf{Varying WD mass}} 
\\
\hline
        
        {}&{}&{}&{}&{}&{}&{}&{}&{}\\
        7 &1.42 &1.25 & 8.0  &138.8 & 306  &   -0.013& +244 & +671\\
        8 &1.42 &1.0  & 8.0  &145.7 & 40   &   -0.013& +233 & +651\\
        9 &1.42 &0.7  & 8.0  &155.7 & 7    &   -0.015& +218 & +628\\
        \hline
    \end{tabular}
    \caption{Model parameters:- The input parameters: initial donor mass ($M_{\rm AGB,ini}$), initial WD mass ($M_{\rm WD,ini}$), initial separation ($a_{\rm ini}$) and corresponding initial orbital period ($P_{\rm orb,ini}$); and output data: total number of nova cycles, total net WD mass change ($\Delta M_{\rm WD}$), total net change in separation ($\Delta a$) and total net change in orbital period ($\Delta P_{\rm orb}$). *Note: shows the hypothetical case with a separation that is below the limit allowed for BHL accretion,  chosen for comparison.}
    \label{tab:1}
\end{table*}

\section{Method of computation and models}\label{main}

\subsection{Method of computation}\label{sec:sec21}
Our simulations were carried out using the  combined binary evolution code \cite []{2020NatAs...4..886H} that utilizes a hydro-static stellar evolution code \cite[]{2009MNRAS.395.1857K} and a nova evolution code \cite[]{1995ApJ...445..789P,2007MNRAS.374.1449E,2015MNRAS.446.1924H}. The hydrostatic stellar evolution code directs the star's evolution from its pre-main sequence phase to the formation of a dense WD, spanning all stellar stages, i.e., pre-main sequence, main sequence, RGB, horizontal branch, AGB, and concluding with a WD. The nova evolution code, a hydrodynamic Lagrangian code, was initially written for modeling multiple consecutive nova eruptions using a given initial WD model and a constant externally given accretion rate \cite[]{1995ApJ...445..789P,2007MNRAS.374.1449E,2015MNRAS.446.1924H}. The two codes were adapted and combined for calculating RLOF in cataclysmic variables (CVs) \cite[]{2020NatAs...4..886H,2021MNRAS.505.3260H} while considering the actual mass transfer rate calculated at each time step, as well as angular momentum loss (AML) from gravitational radiation (GR) and magnetic braking (MB) \cite []{2015ApJS..220...15P}.
In a subsequent modification by \cite{2021MNRAS.501..201H}, this code was adapted to compute the accretion rate in SySs using the BHL prescription requiring the introduction of an additional free parameter --- the binary separation. This is because for CVs where the donor red dwarf (RD) fills its Roche-lobe, the separation is determined by using the stellar masses and the RD radius to initiate the simulation when the RD's radius exactly fills its RL. However, when the case is BHL accretion in a widely separated binary, the separation is a free parameter that needs to be given as initial input along with the binary masses, thus, there are three input parameters required for wind accretion modeling. 
Since a cloud of wind is present in the vicinity of the WD, an AML term was incorporated into the combined code \cite[]{2021MNRAS.501..201H} to accommodate for the effect of drag on the orbit \cite[]{1976ApJ...204..879A}. We have updated our code (described in \cite{2021MNRAS.501..201H} and \cite{2024MNRAS.527.4806V}) to now include an additional AML term that compensates, at each timestep, for mass transfer from the AGB to the WD as well as mass lost from the system --- angular momentum term considering mass transfer (MT).
The code operates by initially considering the given separation and masses of the binary components. It calculates the AML of the system and quantifies AML loss due to mass loss, GR, MB \cite[]{2015ApJS..220...15P}, drag \cite[]{1976ApJ...204..879A} and MT (a correction factor for mass change, obtained by differentiating the orbital momentum without assuming mass conservation) as follows:
\begin{equation} \label{MB}
   \rm \dot{J}_{MB} = -6.82\times10^{34}M_{AGB}R^{4}_{AGB}P^{-3}_{\mathrm{orb}} (\mathrm{dyn\; cm}),\\ 
\end{equation}

\begin{equation} \label{GR}
    \rm \dot{J}_{GR} = - \frac{32}{5c^{5}}  \biggl(\frac{2\pi G}{P_{\mathrm{orb}}}\biggl)^{7/3} \frac{(M_{AGB}M_{WD})^{2}}{(M_{AGB}+M_{WD})^{2/3}},\\ 
\end{equation}

\begin{equation}\label{MT}
\begin{split}
\dot{J}_{\rm MT} &= \dot{M}_w \biggl(\frac{G a}{M_{\rm WD} + M_{\rm AGB}}\biggr)^{1/2} 
\biggl( M_{\rm WD} - \zeta_{\rm BHL} M_{\rm AGB} \\
 &- \frac{M_{\rm WD} M_{\rm AGB}}{2} \cdot 
\frac{1 - \zeta_{\rm BHL}}{M_{\rm WD} + M_{\rm AGB}} \biggr)
\end{split}
\end{equation}

\begin{equation} \label{drag}
   \rm D_w= \pi \rho_w r^2_a v^2_w.
\end{equation}
where $\dot{J}_{\mathrm{MB}}$, $\dot{J}_{\mathrm{GR}}$ and $\dot{J}_{\mathrm{MT}}$ are the change in orbital angular momentum due to MB, GR, MT and ${D}_{\rm w}$ is the drag force respectively. $M_{\mathrm{WD}}$ is the mass of WD, $M_{\mathrm{AGB}}$, mass of the AGB, $R_{\mathrm{AGB}}$, radius of the AGB, $G$ is the gravitational constant, $c$ is the speed of light, $v_{\rm w}$ is the relative velocity of the wind with respect to the WD, $\dot M_{\rm w}$, is the mass loss rate (wind rate) from the AGB obtained from the hydrostatic stellar evolution code \cite[]{2009MNRAS.395.1857K} and $\zeta_{\rm BHL}$, is the fraction of matter that that is transferred to the WD, which is between '0' and '1' and is calculated by dividing the accretion rate by the wind rate from the previous timestep ($\zeta_{\rm BHL}=\dot{M}_{\rm acc}/\dot{M}_{\rm w}$, where $\dot{M}_{\rm acc}$ is the accretion rate).
The total AML of the system is computed for each timestep, which is then utilized to determine the change in the separation of the system for that timestep and the rate of mass accretion onto the WD, all while following the evolution of the rapidly evolving giant, as well as of the WD which produces periodic nova eruption. The change in the separation of the system as a result of the nova eruption is calculated as: 
\begin{equation}\label{dela}
     \Delta a = 2a \biggl(\frac{m_{\mathrm{ej}}-m_{\mathrm{acc}}}{M_{\rm WD}} + \frac{{m_{\rm acc}}}{M_{\rm AGB}}\biggl),
\end{equation}
where $m_{\text{acc}}$ and $m_{\text{ej}}$ are the accreted and ejected masses of the nova cycle. The accretion rate, which was calculated at each timestep following \cite{2024MNRAS.527.4806V} (Equations 2-6), is given as:

\begin{equation}\label{Accretionrate}
    \dot{M}_{{\rm acc}} = \frac{ G^2 M^2_{\rm WD} }{a^{2}v_{\rm w}^4\biggl(\frac{G(M_{\rm WD}+M_{\rm AGB})}{av^2_{\rm w}}+1\biggl)^{3/2}}\dot{M}_{\rm w},
\end{equation}
where `$a$' is the binary separation recalculated at each timestep as described above. The wind velocity at the distance `$a$' can be calculated as:
\begin{equation}\label{windvelocity}
    v_{\rm w}= v_\infty+\frac{R_{\rm AGB}}{a}(v_{\rm s}-v_\infty).
\end{equation}
where ${v_\infty}$ is the terminal wind velocity for which we adopt the value of 20 km $\mathrm{s^{-1}}$ following \cite[]{Kashi_2009,2021MNRAS.501..201H}, and ${v_{\rm s}}$ is the sound velocity at the surface of the AGB. For a more detailed description
we refer to  
\citet[][and references within]{2024MNRAS.527.4806V}.

\subsection{Models}\label{sec:sec23}

We applied our simulation method to different initial binary combinations of WD masses, AGB masses and binary separations. We chose three initial main sequence (MS) masses of 1.5, 2.5 and 3.5 $\rm{M_{\odot}}$ with total evolutionary lifetimes of 2.8$\times 10^{9}$, 7.5$\times 10^{8}$ and 2.9$\times 10^{8}$ years, respectively. These models entered the AGB phase with masses of 1.42, 2.49 and 3.49 $\rm{M_{\odot}}$ respectively, and they reached these evolutionary points after 2.7$\times 10^{9}$, 7.1$\times 10^{8}$ and 2.5$\times 10^{8}$ years for each model respectively. 
These AGB donors were chosen such that they exhibit diverse characteristics---such as different evolutionary time, mass, radius, thermal pulses, wind rates during their AGB phase —to study their effects on different WD masses at varying binary separations and the resulting nova outcomes. The initial simulation point for each AGB donor is chosen such that they initially experience a wind rate of approximately $\rm 10^{-10}~\rm{M_\odot~yr^{-1}}$, and we refer to this point as our simulations' ``$t=0$''. These AGB models were used, and the simulations were carried out using the modified binary evolution code by \citet{2021MNRAS.501..201H}. 
We combined each of the three AGB models with three different initial WD mass models and different initial binary separations, specified in Table \ref{tab:1}. For all our simulations, we ensured the chosen separation to be well within the BHL regime, as determined by \cite{2025ApJ...980..224V}---except one, for which we intentionally used a separation that violates this rule (marked with an asterisk in Table \ref{tab:1}), as an academic test case as explained later in section \ref{32}. Each simulation was carried out until the donor shed its entire envelope and the wind became negligible, marking the end of the AGB phase.

\begin{figure*}
    \centering
    \includegraphics[trim={2.0cm 2.0cm 2.0cm 1.0cm},clip,width=2\columnwidth]{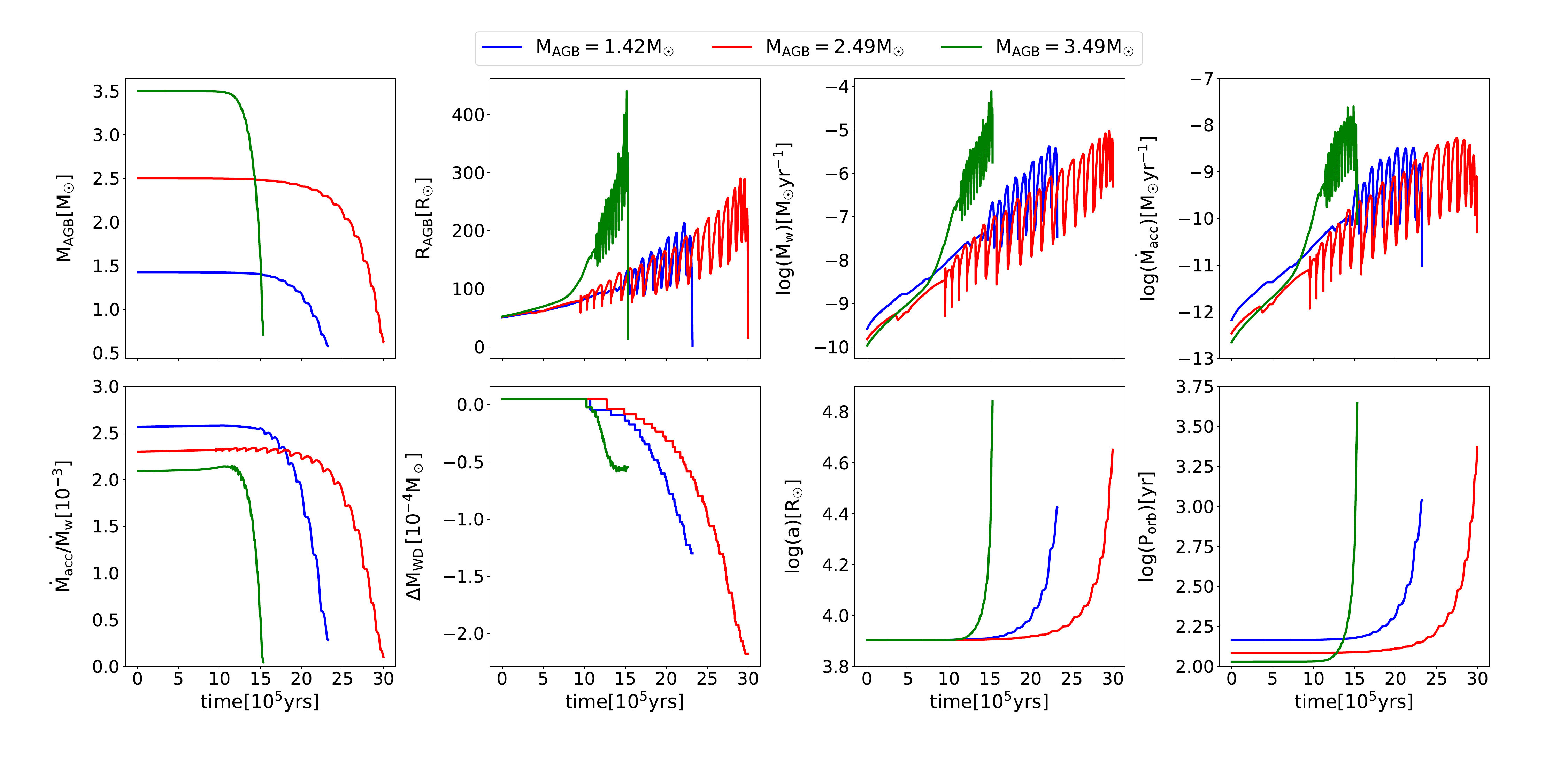}
    \caption{AGB mass ($M_{\rm AGB}$), its radius ($R_{\rm AGB}$) and its wind rate ($\dot{M_{\rm w}}$), the mass transfer efficiency ($\dot{M}_{\rm acc}/\dot{M}_{\rm w}$), the change in WD mass ($\Delta M_{\rm WD}$), the binary separation ($a$), and the orbital period ($P_{\rm orb}$), for an initial 1.0$M_\odot$ WD with an initial 8000$\rm{R_\odot}$ binary separation, and three different initial AGB masses: 1.42 (blue), 2.49 (red) and 3.49$ {M_\odot}$ (green). The final AGB masses are 0.58, 0.62 and 0.71$M_\odot$ respectively. The WD shows an overall mass loss regardless of the choice of initial AGB mass, and the orbital separation and period increase for all three models, as most of the matter is lost from the system. }
    \label{Fig1}
\end{figure*}

During the course of a star's evolution on the AGB, it experiences thermal pulses that cause rapid contractions and expansions (e.g., \citealt{1983A&A...127...73B,1993ApJ...413..641V,dorfi1998agb}), leading to substantial changes in both the radius and the wind rate of the star. Figure \ref{Fig1} illustrates that the three AGB stars that we use as our donors in this work, exhibit significantly varying radii as a result of thermal pulses, with more bloated radii producing higher wind rates. Since stellar evolution dictates that the wind rate is not necessarily monotonically correlated with AGB mass (e.g., \citealt{2014ApJ...790...22R,2015ASPC..497..229M}), the evolutionary lifetime of a star in the AGB phase is not necessarily monotonically correlated with the AGB mass, as we see for our donors in Figure \ref{Fig1}.

In this work we track the evolutionary changes in the masses of the WD and the AGB donor, its wind rate, and the systems separation and orbital period, while monitoring and incorporating the accretion rate ($\dot{M}_{\rm acc}$) at each timestep. We investigate the variations of mass transfer efficiency $\mathrm{\zeta}_{BHL}=\mathrm{\dot{M}_{acc}/\dot{M}_w}$. We calculate the mass retention efficiency, $\eta=(m_{\rm acc}-m_{\rm ej})/m_{\rm acc}$, the average accretion rate ($\dot{M}_{av,acc}$), the maximum temperature ($T_{\mathrm{max}}$) attained by the WDs during each eruption and we also follow their core temperatures ($T_{\mathrm{c}}$) at each timestep throughout evolution. We also record the hydrogen ($X_{\rm ej}$), helium ($Y_{\mathrm{ej}}$), and heavy element ($Z_{\mathrm{ej}}$) abundances in the ejecta, per eruption and track the change in orbital separation. 

\begin{figure*}
    \centering
    \includegraphics[trim={2.0cm 2.0cm 2.0cm 1.0cm},clip,width=2\columnwidth]{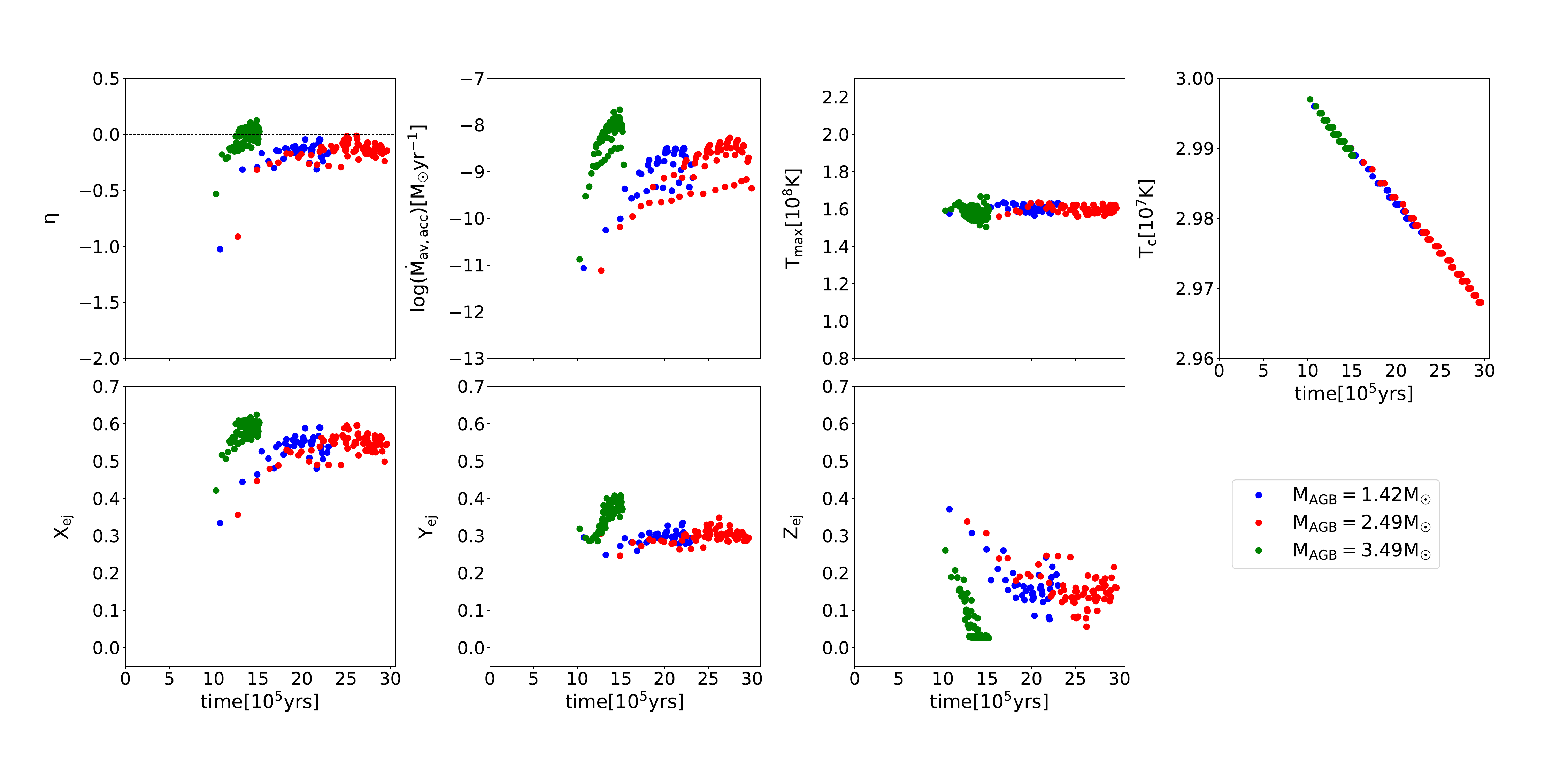}
    \caption{Mass retention efficiency ($\eta$), average accretion rate ($\dot{M}_{av,acc}$), maximum temperature ($T_{\rm max}$), core temperature ($T_{\rm c}$) and mass fractions of hydrogen ($X_{\rm ej}$), helium ($Y_{\rm ej}$), and heavy elements ($Z_{\rm ej}$) in the ejecta per cycle. Models with higher accretion rates exhibit greater WD mass retention and shorter core cooling times. This leads to a higher hydrogen fraction and helium content, and lower fraction of heavier elements in the ejecta.} 
    \label{Fig2}
\end{figure*}

\begin{figure*}
\centering
    \includegraphics[trim={2.0cm 2.0cm 2.0cm 1.0cm},clip,width=2\columnwidth]{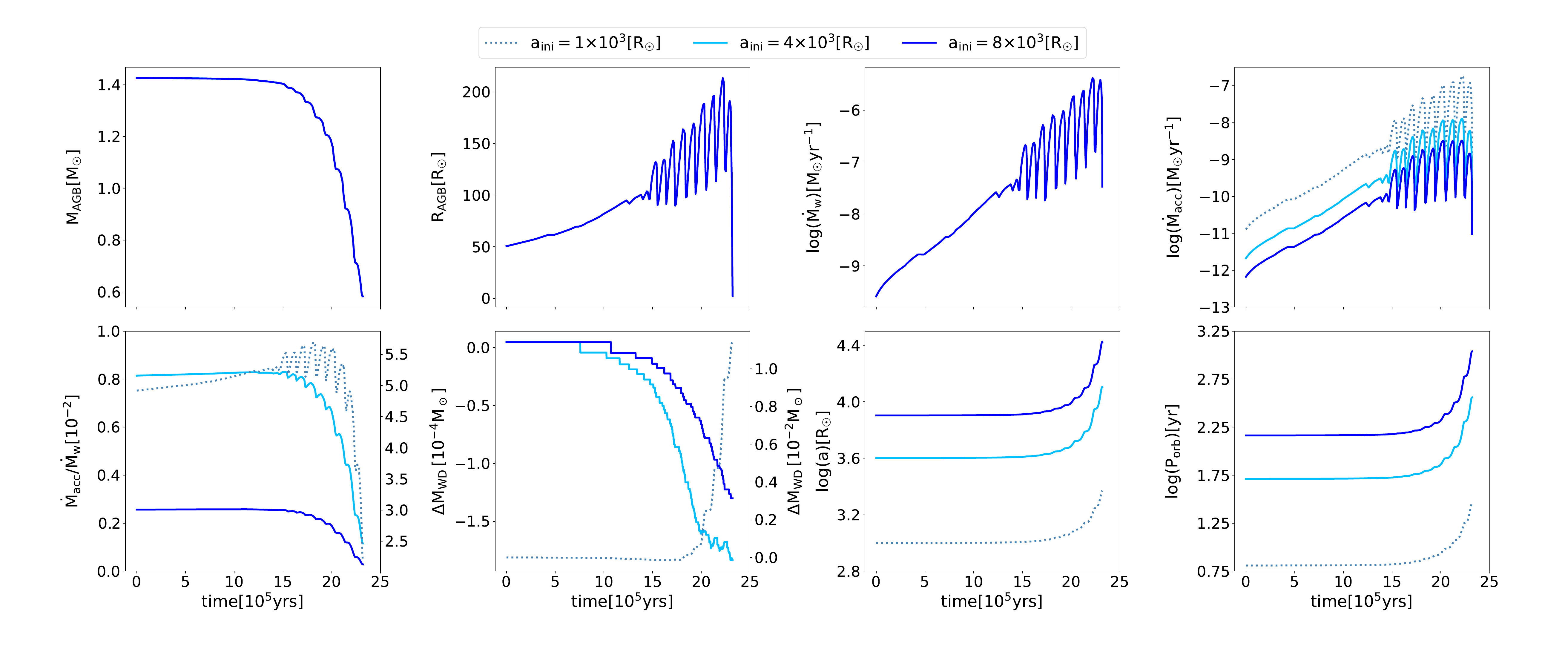}
    \caption{Description as in Figure \ref{Fig1}, for an initial $M_{\rm WD} = 1.0M_\odot$ and $M_{\rm AGB} = 1.42M_\odot$, with three initial separations of $a_{\rm ini}$ = 1000 (dotted blue), 4000 (solid sky-blue) and 8000$\rm{R_\odot}$(solid blue). The panel showing ${\dot{M}_{\rm acc}/\dot{M}_{\rm w}}$ has two y-axes: the left y-axis corresponds to $a_{\rm ini} = 4000$ and $8000\rm{R_\odot}$, and the right y-axis corresponds to $a_{\rm ini} = 1000\rm{R_\odot}$. The same applies to the panel of $\Delta M_{\rm WD}$. The model with an initial separation of $a_{\mathrm{ini}} = 1000\rm{R_\odot}$ shows an increase in WD mass due to a higher accretion rate, resulting in reduced mass ejection during nova eruptions. The separation and orbital period continue to increase as most of the mass is lost from the system.}
    \label{Fig3}
\end{figure*}

\begin{figure*}
    \centering
    \includegraphics[trim={1.0cm 2.0cm 2.0cm 1.0cm},clip,width=2.0\columnwidth]{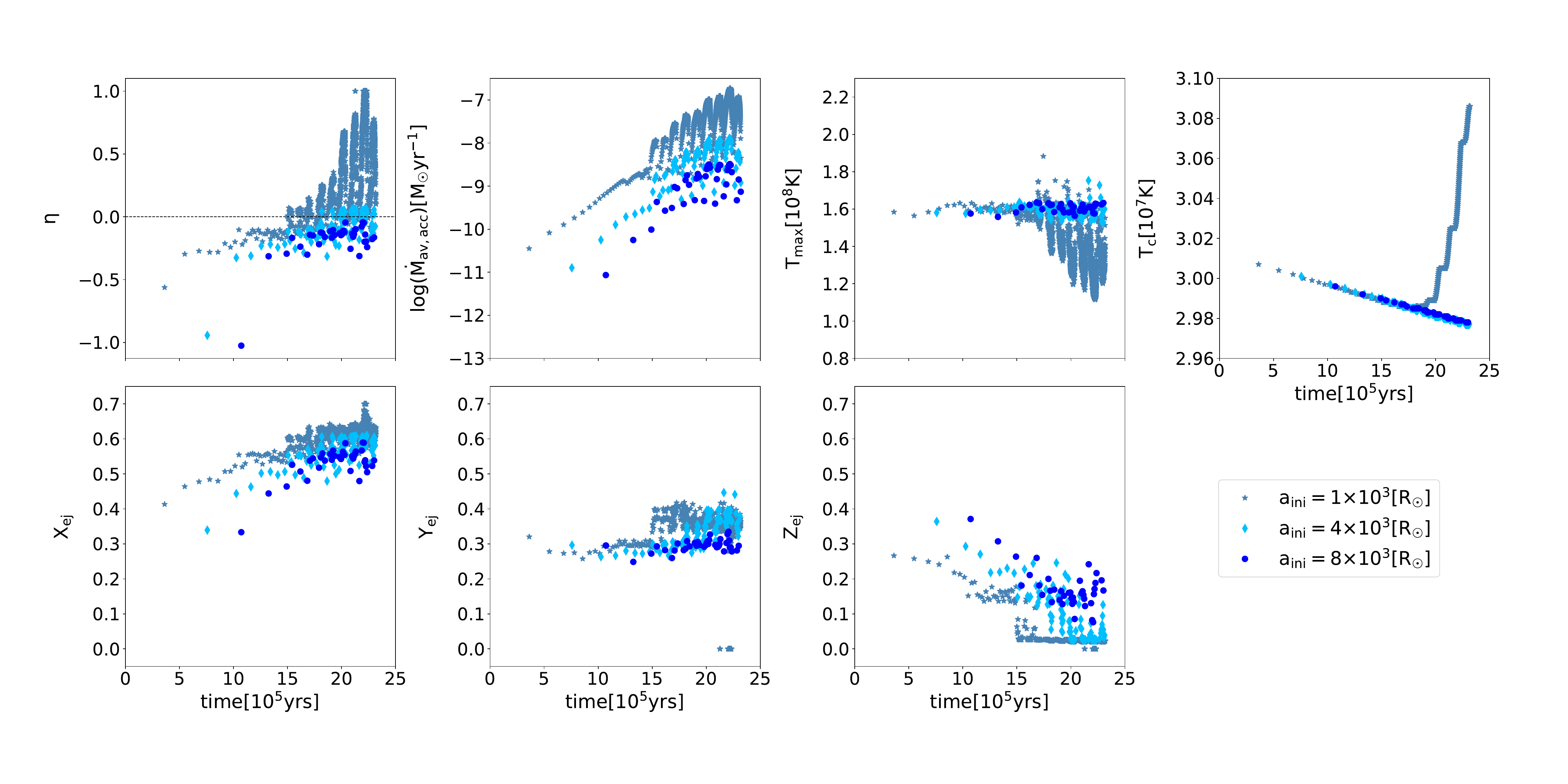}
    \caption{Description as in Figure \ref{Fig2}, for an initial $M_{\rm WD} = 1.0M_\odot$ and $M_{\rm AGB} = 1.42M_\odot$, with three initial separations of $a_{\rm ini}$ = 1000, 4000, 8000$\rm{R_\odot}$. }
    \label{Fig4}
\end{figure*}

\begin{figure*}
    \centering
    \includegraphics[trim={2.0cm 2.0cm 2.0cm 1.0cm},clip,width=2\columnwidth]{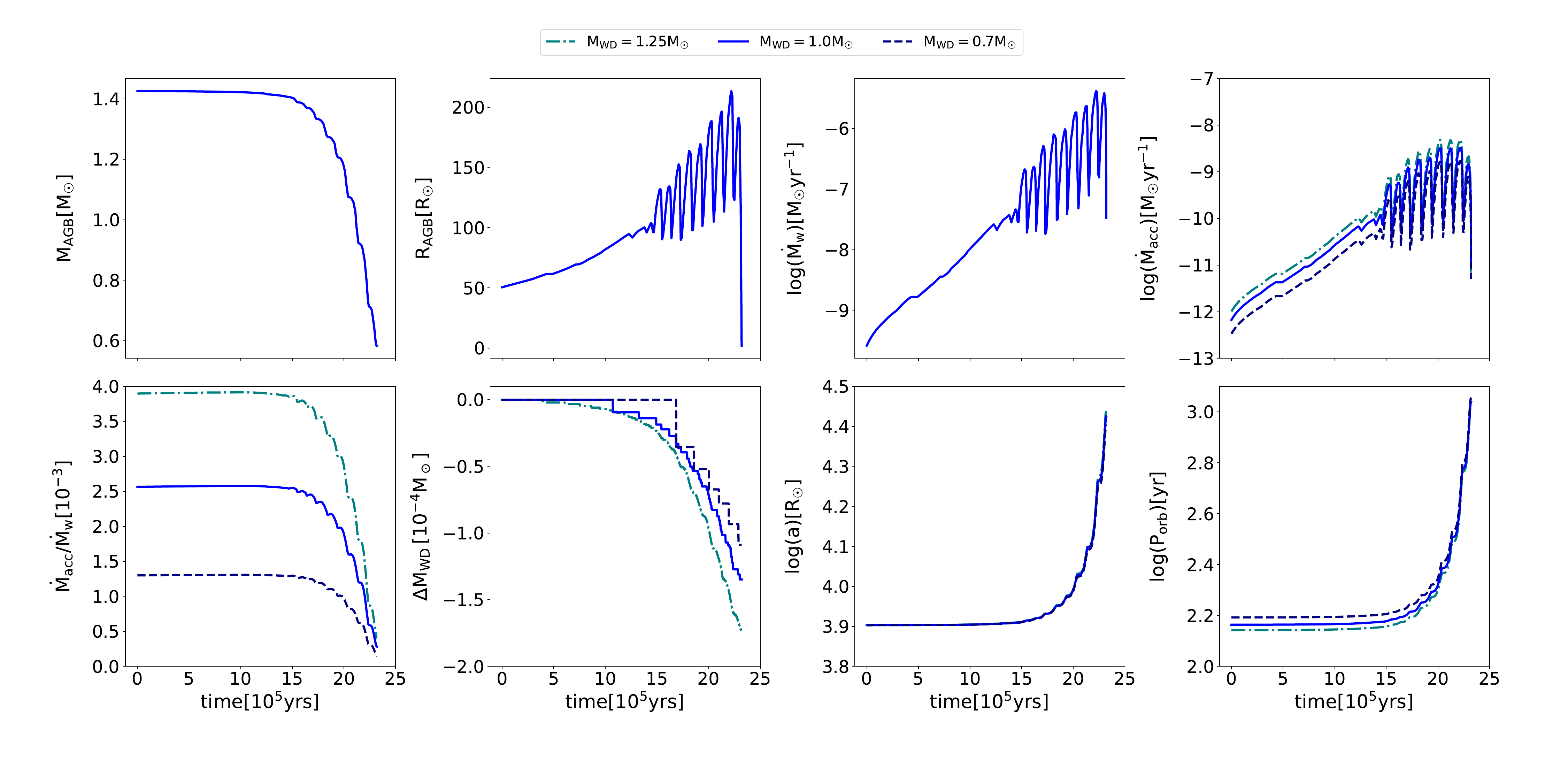}
    \caption{Description as in Figure \ref{Fig1}, for an initial $M_{\rm AGB} = 1.42\rm{M_{\odot}}$ and an initial separation of $a_{\rm ini} = 8000 \rm{R_\odot}$, with three different WD masses: $M_{\rm WD} = 1.25$ (dash-dotted teal), $1.0$ (solid blue), and $0.7 \rm{M_\odot}$ (dashed navy-blue). Models with more massive WDs lose more mass due to higher accretion rates and a higher number of eruptions, causing increased mass loss from the system. The orbital separation and period follow the general trend of increasing values.} 
    \label{Fig5}
\end{figure*}

\begin{figure*}
    \centering
    \includegraphics[trim={2.0cm 2.0cm 2.0cm 1.0cm},clip,width=2\columnwidth]{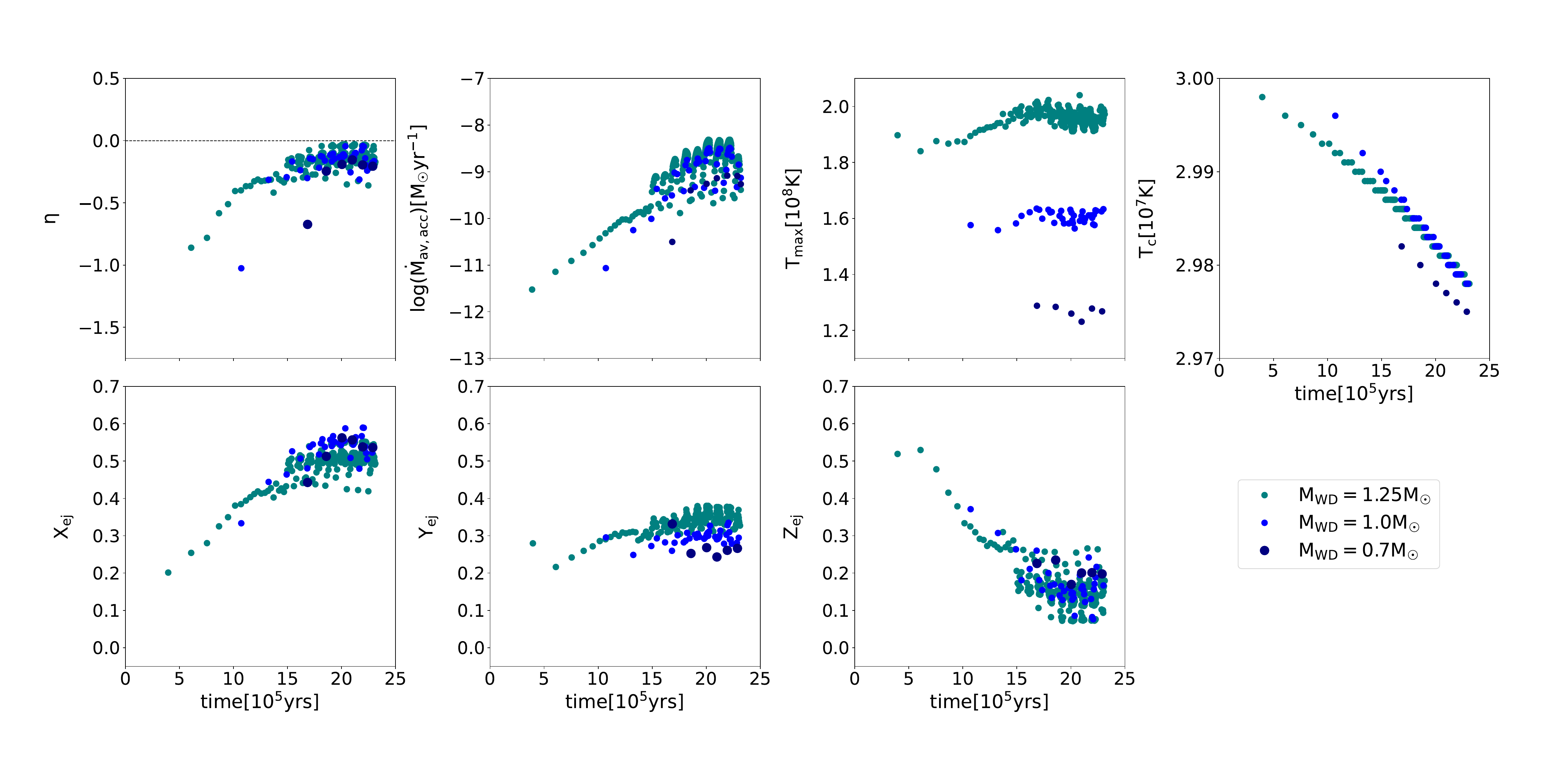}
    \caption{Description as in Figure \ref{Fig2}, for an initial $M_{\rm AGB} = 1.42M_\odot$ and an initial separation of $a_{\rm ini} = 8000\rm{R_\odot}$, with three different WD masses: $M_{\rm WD} = 1.25$, $1.0$, and $0.7 \rm{M_\odot}$.}
    \label{Fig6}
\end{figure*}

\begin{figure}
    \hspace{-0.5cm}
    \includegraphics[trim={2.0cm 5.0cm 2.0cm 2.0cm},clip,width=0.5\textwidth]{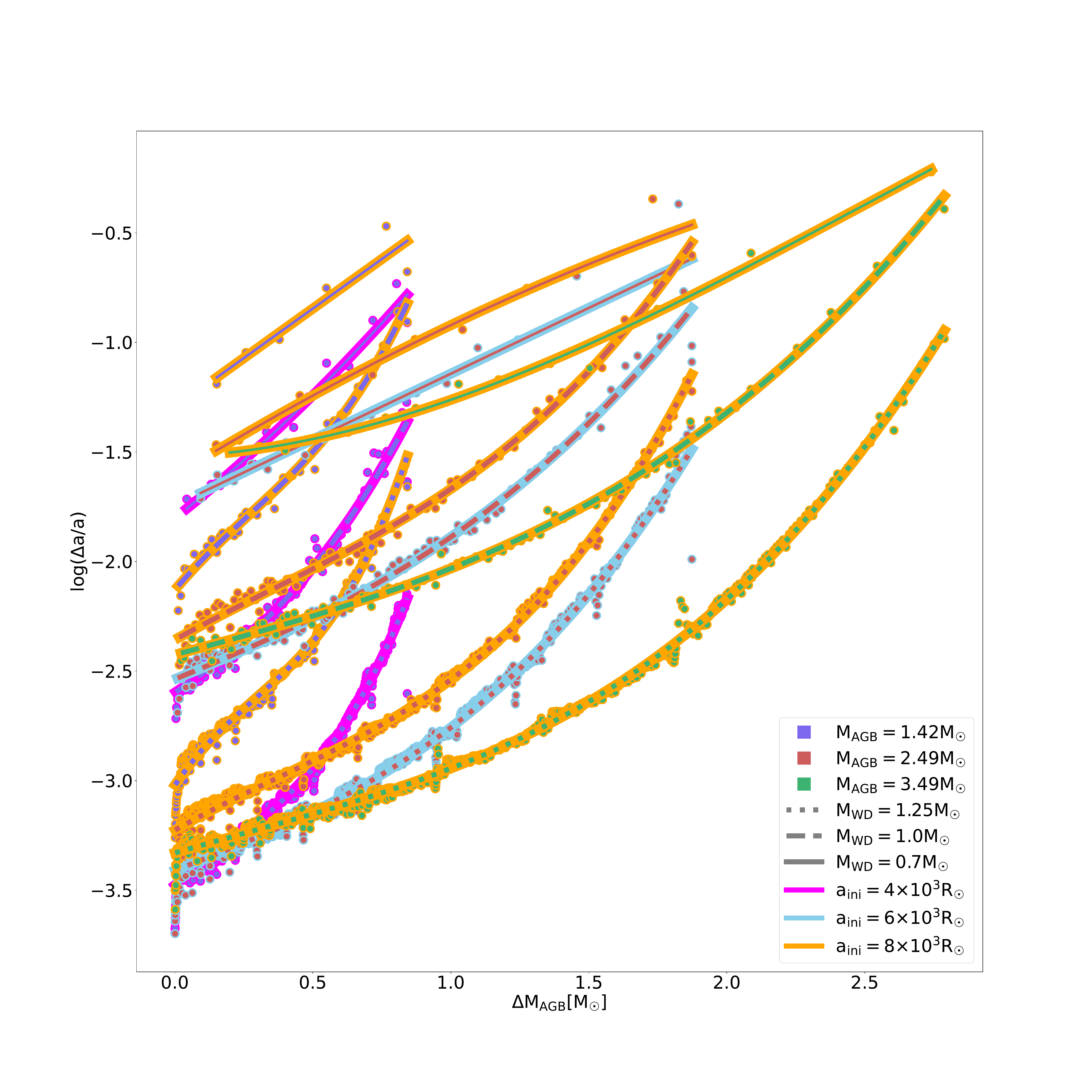}
    \caption{The relative change in orbital separation, $\Delta a/a$, (on a logarithmic scale) vs. the change in donor mass, per cycle for all our models. An eye-fit line, based on the data, is shown to illustrate the overall trend in orbital evolution. The change in separation increases with: more mass loss; larger initial separations; and more massive WDs.}
    \label{fig:Fig7}
\end{figure}

\begin{figure}
\centering
    \includegraphics[trim={2.0cm 1.0cm 2.0cm 1.0cm},clip,width=0.5\textwidth]{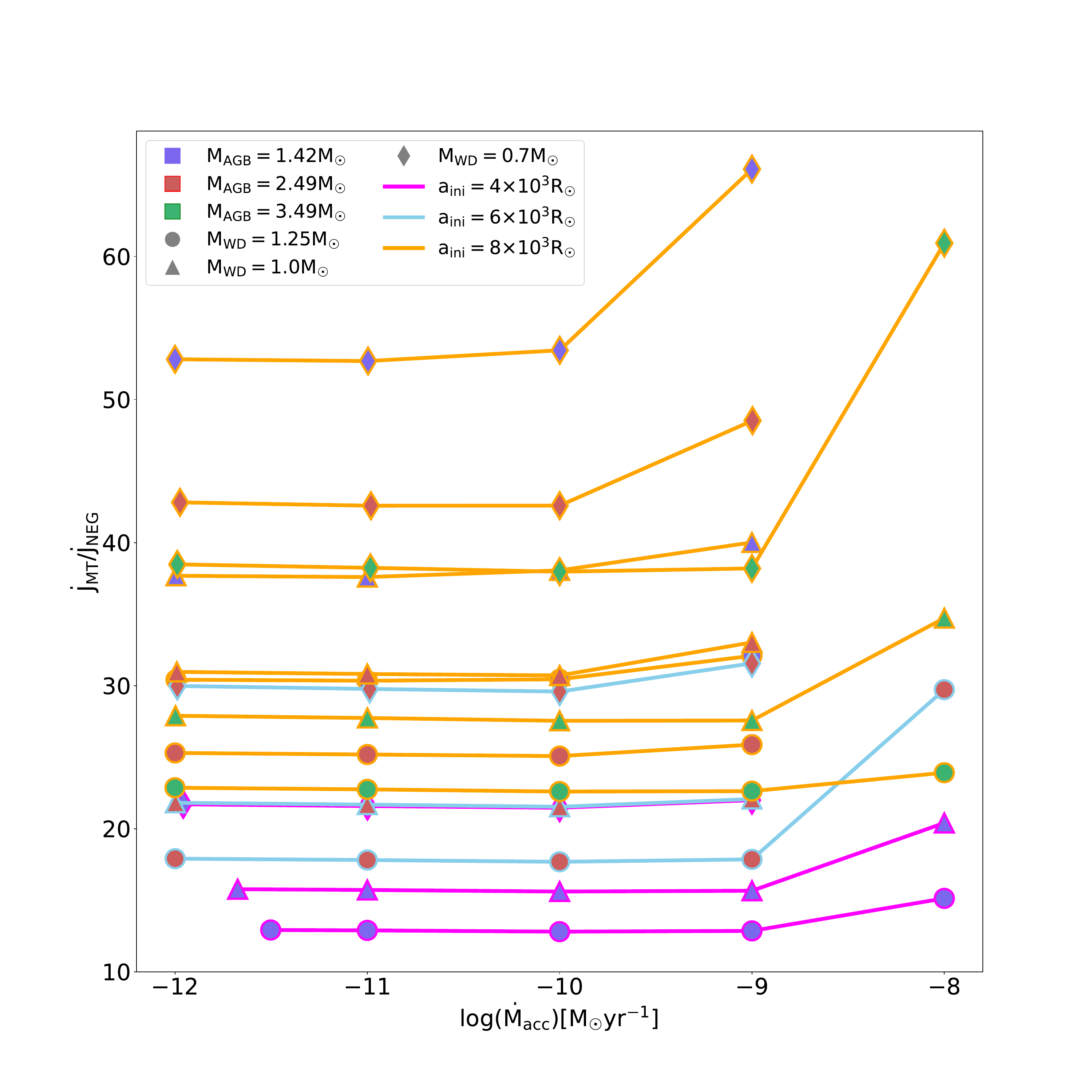}
    \caption{Ratio of the component of the mass transfer angular momentum to the sum of the other components $\dot{J}_{MT}/\dot{J}_{NEG}$ vs. different accretion rates during the evolution of each model. The ratio increases with increase in accretion rate, higher separation and less massive WDs.}
    \label{fig:Fig8}
\end{figure}

\begin{figure}
\centering
    \includegraphics[trim={1.0cm 1.0cm 2.0cm 1.0cm},clip,width=0.5\textwidth]{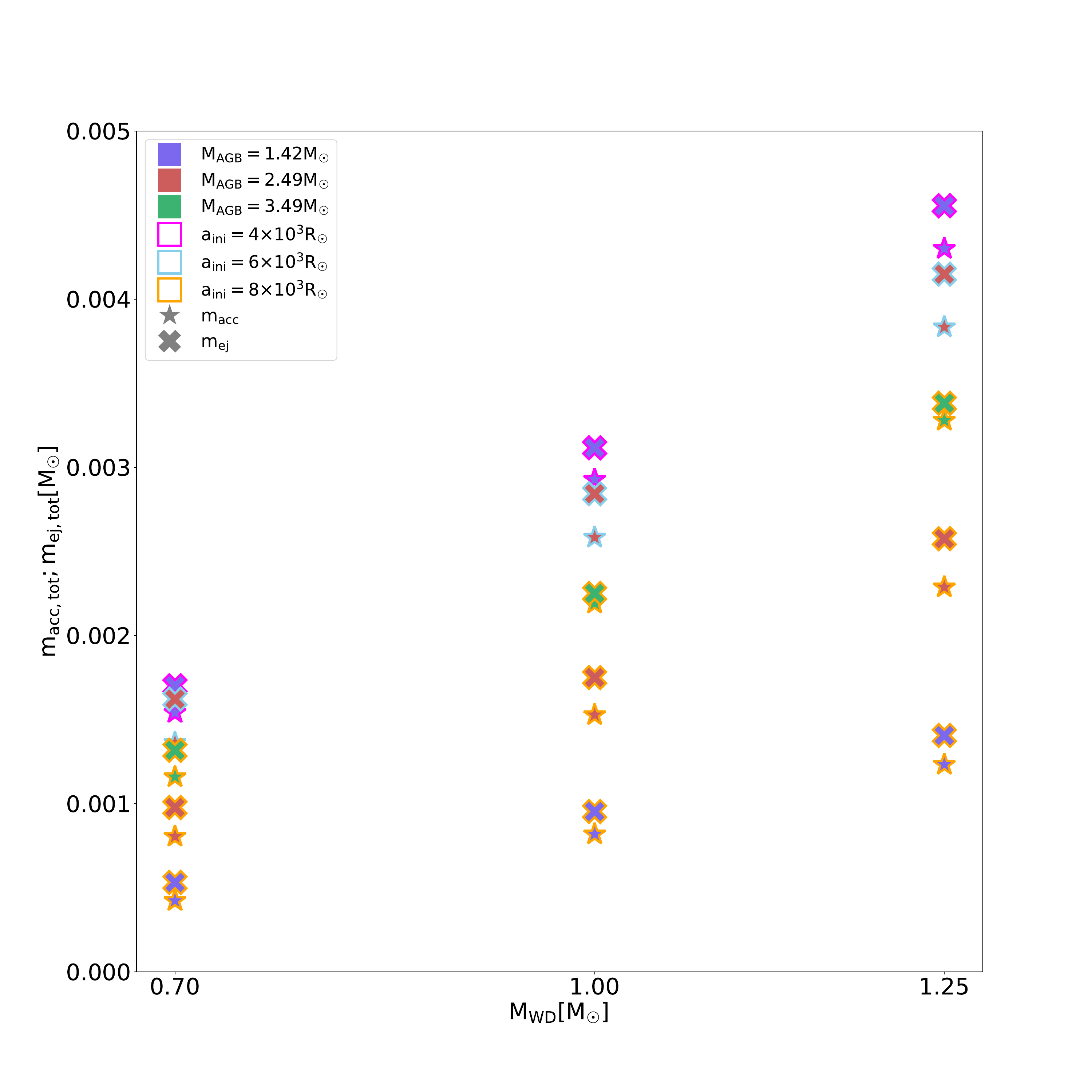}
    \caption{Total accreted and ejected masses ($m_{\rm acc,tot}$ and $m_{\rm ej,tot}$ respectively) throughout the AGBs' lifetimes for each WD mass across all models.  All models has higher mass ejected than mass accreted. More massive WDs exhibit both higher accreted and lost mass due to their higher accretion rates. This results in overall greater retention during the accretion phase, but also increased mass loss driven by a higher number of eruptions.  }
    \label{fig:Fig9}
\end{figure}

\section{Results}\label{sec:results}

The work carried out by \cite{2024MNRAS.527.4806V} explored the impact of separation and WD mass on nova eruptions in SySs with a 1.0 $\rm  M_\odot$ AGB donor. This study is an expansion to it by additionally investigating how different AGB donors affect the system's eruptive behavior. For better understanding, we have grouped our results into three subsections. The first subsection (\S\ref{sec:vary_AGB}) explores the impact of different AGB masses while keeping the initial WD mass and binary separation fixed. The second subsection (\S\ref{32}) compares different binary separations with the same initial AGB and WD masses. The third subsection (\S\ref{33}) varies the WD mass, keeping the initial AGB mass and binary separation identical. 

\subsection{Varying  AGB mass}\label{sec:vary_AGB}

We have taken three different AGB models with masses of 1.42, 2.49 and 3.49$\rm{M_{\odot}}$, each paired with a 1.0$\rm{M_{\odot}}$ WD model. All the three simulations were given an initial separation of 8000$\rm{R_\odot}$.

Figure \ref{Fig1} shows the evolution of different characteristics of the binary systems, namely, the mass of the AGB ($M_{\rm AGB}$), its radius ($R_{\rm AGB}$) and its wind rate ($\dot{M_{\rm w}}$), the accretion rate ($\dot{M}_{\rm acc}$), the mass transfer efficiency (${\dot{M}_{\rm acc}/\dot{M}_{\rm w}}$) which is the fraction of matter (or portion of wind) that gets transferred to the WD, the change in mass of the WD ($\Delta M_{\rm WD}$), the binary separation ($a$), and the orbital period ($P_{\rm orb}$) of the binary system. The figure shows that for all three simulations in this set, the WD consistently loses mass, of order, $\sim$0.01\% throughout the evolution.
This is because the mass transfer efficiency is very low---of order $<$0.27\% (as shown in Figure \ref{Fig1}), which leads to average accretion rates of order $10^{-12}-10^{-7.5}~\rm{M_{\odot}~yr^{-1}}$ that are almost entirely within the nova regime of net mass loss \cite[]{2005ApJ...623..398Y}. The lower accretion rate leads to the accumulation of matter onto the WD surface over a longer timescale, enabling the accreted material to diffuse and mix sufficiently with the WD surface. As a result, a nova eruption will lead to the ejection of material from deep inside the WD surface, carrying away a portion of WD matter in addition to the accreted material (e.g., \citealt{1994ApJ...424..319K,1995ApJ...445..789P,2005ApJ...623..398Y,2012BaltA..21...76S,2015MNRAS.446.1924H,2021MNRAS.505.3260H,2024MNRAS.527.4806V}). 
The separation for these three models show a change of $\sim\times 3-10$ over evolution and the orbital period shows a change of $\sim\times10-50$. This increase in separation reflects the pattern of the low accretion efficiency expressing that most of the wind from the AGB is lost from the system, causing the separation to increase, and as it increases, the efficiency decreases --- a positive feedback process --- leading to a runaway separation growth. This is because the AML due to MT dominates over GR, MB, and drag forces, thus acting as the dominant AML mechanism. As we see from the accretion efficiency in Figure \ref{Fig1}, almost all the mass that is blown in the AGB's wind is expelled from the system, which is the reason why the MT component of the total AML is dominant. However, if the mass transfer were efficient, as, say in RLOF cases in cataclysmic variables, this MT component would be significantly smaller, allowing for the other AML components (such as GR and MB) to dominate, thus resulting in decreasing separation and orbital period \cite[]{2020NatAs...4..886H,2021MNRAS.505.3260H}.

The evolution of selected eruption properties, namely, the mass retention efficiency ($\eta$, defined as $(m_{\rm acc}-m_{\rm ej})/m_{\rm acc}$), the average accretion rate ($\dot{M}_{av,acc}$), the maximum temperature attained during an eruption ($T_{\rm max}$), the WD's core temperature ($T_{\rm c}$), and the mass fractions of hydrogen ($X_{\rm ej}$), helium ($Y_{\rm ej}$), and heavy elements ($Z_{\rm ej}$) in the ejecta per eruption are shown in Figure \ref{Fig2}. We find that higher wind rates, as they directly cause higher accretion rates, support better mass retention after each eruption, thus, the 3.49$M_\odot$ AGB model (green) shows the highest mass retention efficiency --- including a few cycles with positive values of $\eta$ --- for cycles with higher average accretion rates $\sim 10^{-8.5}-10^{-7.5}~\rm{M_{\odot}~yr^{-1}}$, high enough to lead to nova eruptions with a net mass increase. We do not see a positive mass retention at any point for the other two AGB donors since their average accretion rates were always $< 10^{-8.5}~\rm{M_{\odot}~yr^{-1}}$. The values of $T_{\rm max}$ for all models remain approximately in the same range due to the overall low accretion rates leading to similar outcomes. The small fluctuations are due to changes in wind rate, which are reflected in the accretion rate, slightly increasing or decreasing the temperature from cycle to cycle. The WD's core temperature ($T_c$), which shows a consistent decrease, is again correlated with the low accretion rates operating at these evolutionary phases. This results in long intervals between eruptions, allowing enough time for the WD core to cool before the next eruption, and thereby consistently reducing the core temperature from one eruption to the next, which is consistent with previous nova models  \cite[]{2005ApJ...623..398Y}. 
Since the 3.49$M_\odot$ AGB model produced the highest accretion rate of the three, there was less time for mixing between the accreted matter and the WD material, resulting in a higher amount of hydrogen and helium and a lower fraction of heavy elements in the ejected matter of the nova eruptions, as seen in previous works \cite[e.g.,][]{1995ApJ...445..789P,2005ApJ...623..398Y,2012BASI...40..419S,2020ApJ...895...70S,2019ApJ...879L...5H,2022MNRAS.511.5570H}.

Similar plots for the same three AGB masses with WD masses of 1.25 and 0.7$\rm{M_{\odot}}$, while adopting the same initial binary separation are given in Appendix \ref{A1}, demonstrating that the general behavior of all the parameters shown here remains the same for different WD masses, establishing the robustness of our results.

\subsection{Varying separation}\label{32}

These simulations were carried out with WD and AGB models of 1.0$\rm{M_{\odot}}$ and 1.42$\rm{M_{\odot}}$, respectively, with three different separations---1000, 4000, 8000$\rm{R_{\odot}}$. The smallest separation here is beyond the limit allowed for BHL accretion as determined by \cite{2025ApJ...980..224V}, however, we included it to better understand the trends rising from varying the separation and denote this as a purely hypothetical academic case.
Figure \ref{Fig3} shows the same parameters as given in Figure \ref{Fig1} and Figure \ref{Fig4} shows the same parameters as shown in Figure \ref{Fig2}. For the initial separations of 4000 and 8000$R_\odot$ we see a consistent decrease in the WD's mass, with the mass loss being \textit{higher} for smaller separations, even though the accretion rate and mass retention efficiency are higher for smaller separation (as seen in Figure \ref{Fig3} and \ref{Fig4}, respectively). So why does the WD in the system with a narrower separation lose mass faster?  This is because the higher accretion rate triggers more frequent nova eruptions, and even though the net mass retention is higher for these cases, it is still not high enough to produce net-mass-gain eruptions (i.e., positive values of $\eta$), thus more eruptions leads to more erosion of the WD. In contrast with these two models, the 1000$R_\odot$ separation shows the WD mass to increase over time. This is because this separation leads to a more efficient mass transfer rate, yielding accretion rates that are high enough to produce novae with net mass gain (positive values of $\eta$, see Figure \ref{Fig4}) for a significant time of the evolution, thus, the WD mass increases. We stress, however, that this narrow orbit is purely hypothetical for BHL accretion, based on previous work \cite[]{2025ApJ...980..224V}.
Additionally shown in Figure \ref{Fig3} are the change in separation and orbital period, which in contrast with the previous section, the change in separation here is only $\sim\times2-3$ and in orbital period $\sim\times5-10$, while the more significant change is for wider initial separations, causing further widening.

The more drastic change in separation in the previous section indicates that the initial AGB mass has a stronger influence on the separation change than the initial separation. This is because of the \textit{total amount of mass lost from the system}  leads to a separation increase. This total amount of mass is higher for more massive AGB donors.

Figure \ref{Fig4} shows $\sim$consistent $T_{\rm max}$, for the 4000 and 8000$\rm{R_\odot}$ models, as well as a monotonically decreasing $T_{\rm c}$, while the 1000$R_\odot$ model exhibits cases of lower $T_{\rm max}$ and a rising $T_{\rm c}$ for cycles that attained high accretion rates. The mass fractions in the ejecta ($X_{\rm ej}$, $Y_{\rm ej}$ and $Z_{\rm ej}$) show similar trends as in the previous section because, ultimately, they depend on the accretion rate which is correlated with the separation and wind rate. 
In the previous section, the higher wind rates lead to the higher accretion rates, while in this section it is the closer separation that leads to it. 
The higher accretion rate means that the WD attains the triggering mass in less time, preventing deeper mixing of the accreted matter into the outer layers of the WD core, causing the TNR to occur at a shallower point, so the ejection requires less energy to lift the overlying material from the WD, resulting in fusion for a shorter period, with lower temperatures, until enough energy is generated to expel the ejecta. 
However, it is important to note that even the lowest separation, with its higher accretion rate, did not cause significant changes to the compositions. The same holds true for the other models, indicating that WD-AGB binaries in BHL systems, and their corresponding accretion rates, are not capable of contributing a significant amount of enriched elements to the surroundings as a result of nova eruptions. Additional plots showing models with two different initial binary separations for the 2.49$\rm{M_{\odot}}$ AGB model are provided in Appendix \ref{B1}, still exhibiting similar trends. 

\subsection{Varying WD mass}\label{33}

 To study the effect of the WD mass on the behavior of the system, we show in Figure \ref{Fig5}  three models with the same initial AGB mass of 1.42$\rm{M_{\odot}}$ and an initial binary separation of 8000$\rm{R_{\odot}}$, for three initial WD masses: 0.7, 1.0, and 1.25$\rm{M_{\odot}}$. We find the mass transfer efficiency for the more massive WD ($1.25M_\odot$) to be $\sim\times3$ more efficient than for the lower WD mass ($0.7M_\odot$), however, since the average accretion rate is always $\lesssim10^{-8}M_\odot\rm yr^{-1}$ (Figure \ref{Fig6}), the WD masses decrease throughout evolution. In the long run, the more massive WD lost more mass than the less massive WD because it underwent more eruptions. This is because a more massive WD requires less accreted mass to trigger a TNR, thus, for the given initial AGB donor and separation, a more massive WD experiences more nova eruptions (see Table \ref{tab:1}, and Figure \ref{Fig6}, where each data point represents an eruption). However, when we examine the percentage change in the WD mass (Table \ref{tab:1}), we find that all three models show a $\sim$ similar change, with the 0.7$\rm{M_{\odot}}$ WD exhibiting a slightly higher percentage increase. This is because, although the change in mass is higher for the 1.25 $\rm{M_{\odot}}$ WD, the relative (percentage) change is smaller due to its larger initial mass. In contrast, the 0.7$\rm{M_{\odot}}$ WD loses a smaller \textit{amount} of mass, but when considering the overall change, this model shows the highest \textit{percentage} increase because its initial mass was lower. As a result, the change in mass represents a larger fraction relative to its initial value, but this represents a larger fraction of its initial mass. If the accretion rates were the same across all three models, we would expect more massive WDs to have lower retention efficiencies, as the triggering mass for a nova is attained at a faster rate. However, since massive WDs actually experience higher accretion rates (Equation \ref{Accretionrate}), the overall retention efficiency is shifted upward. This shift allows massive WDs to retain more accreted material, resulting in overall retention efficiencies that are similar across the three models despite their differences in WD mass.

Figure \ref{Fig5} also shows that the WD mass has very little effect on the binary separation (and orbital period). This is because all three models share the same AGB donor, which dictates a common wind (mass-loss) rate, and they also begin with the same initial orbital separation. The only varying parameter is the WD mass. A higher WD mass increases the mass transfer efficiency (see Equation \ref{Accretionrate}). In our models, the average accretion rate ranges from $10^{-11}$ to $10^{-9}~ \rm{M_\odot~{yr}^{-1}}$ for the 0.7$\rm{M_{\odot}}$ WD, and from $10^{-11.5}$ to $10^{-8.25}~ \rm{M_\odot~{yr}^{-1}}$ for the 1.25$ \rm{M_{\odot}}$ WD. However, this has little impact on the orbital separation and period, as the fraction of mass lost remains similarly high in all cases. Orbital evolution is due to change in mass and angular momentum. Since both the mass-loss rate and initial separation are same across models, so is the separation increase. The similarity in the evolution of orbital separation (the shape of the curve, which signifies the rate of change) in Figure \ref{Fig3} arises from the models having the same initial AGB mass as well, while the shift along the vertical axis is simply the different initial separation.

Figure \ref{Fig6} shows that a lower $T_{\rm max}$ is obtained for lower WD masses, as their weaker gravity leads to less compression of the accreted material, resulting in lower peak temperatures during eruptions, as obtained in previous nova evolution works (e.g., \citealt{2005ApJ...623..398Y,2021MNRAS.505.3260H,2024MNRAS.527.4806V}). For a given WD mass, we obtain a fairly constant $T_{\rm max}$ as seen for the low accretion rates in the previous sections.
The core temperatures for all WDs are roughly the same, as they all have sufficient time to cool between eruptions, thus reducing their core temperatures further in agreement with the nova models by \cite{2005ApJ...623..398Y}.
Since more massive WDs undergo more rapid eruptions, the reduced mixing time leads to a higher mass fraction of hydrogen, a lower fraction of helium, and an approximately similar or slightly lower fraction of heavy elements in the ejecta, consistent with the findings of \cite{2021MNRAS.505.3260H} for the 1.25$\rm{M_{\odot}}$ WD. Additional plots with the same initial binary separation, for AGB masses of 2.49$\rm{M_{\odot}}$ and 3.49$\rm{M_{\odot}}$ are provided in Appendix \ref{C1} showing similar trends.

\section{Discussion}\label{sec:Discussion}

\subsection{Why does the binary separation consistently increase?}

Figure \ref{fig:Fig7} shows the relative change in orbital separation ($\Delta a/a$) as a function of the decrease in AGB mass ($\Delta M_{\rm AGB}$), sampled once per cycle for all of our models. 
The figure reveals a number of clear trends. The most prominent one being that as $\Delta M_{\mathrm{\rm AGB}}$ increases, i.e., as the AGB mass decreases, $\Delta a/a$ increases. This rate of change is enhanced with time because as these systems evolve, the separation increases, which enhances the increase in separation. This is because a larger separation yields a smaller $\zeta_{BHL}$, meaning more mass lost,   thus causing more rapid widening of the orbit (Equation \ref{MT}).

An additional trend is revealed by following models of the same initial orbital separation and WD mass,  but different AGB masses. For instance, the orange-bordered lines ($a_{\rm ini}=8\times10^3\rm{R_\odot}$) with a solid type inner line ($M_{\rm WD}=0.7\rm{M_\odot}$), that is blue, red and green ($M_{\rm AGB,ini}=1.42$, 2.49 and 3.49$\rm{M_\odot}$ respectively), this set of three models shows that an initially more massive AGB begins with a lower initial rate of orbital change (i.e., when total mass loss is still low), but ends with the largest overall increase in $\Delta a/a$. The difference in the initial $\Delta a/a$ is directly related to the mass loss rate --- the $1.42M_\odot$ AGB begins with the highest wind rate (see Figure \ref{Fig1}), thus it has the most rapid separation change. {The reason the more massive AGB ($3.49\rm{M_\odot}$) leads to the largest overall change in separation is that it has more mass to lose over the course of its evolution.

Another trend that is apparent from the figure may be seen by following models with the same initial AGB mass and WD mass but different separations. For instance, red-dashed inner lines ($M_{\rm AGB,ini}=2.49M_\odot$; $M_{\rm WD}=1.0M_\odot$) with orange and light-blue line borders ($a=8\times10^3$ and $6\times10^3\rm{R_\odot}$ respectively). By examining these curves, we find that the increase rate of $\Delta a/a$ behaves the same even though $a_{\rm ini}$ is different, i.e., both curves show $\Delta a/a$ to increase by a total of $\sim1.5$ orders of magnitude --- this correlation is evident by the lines being parallel. Following these two curves also shows that the initially wider orbit leads to higher initial and higher final values of $\Delta a/a$ --- meaning that the separation changes faster for wider orbits. This is because a higher separation leads to a lower accretion rate, which means more mass is lost from the system, leading to  further increase of separation.

We see another $\Delta a/a$ trend for models with the same initial AGB masses and separation, for instance, pink-bordered lines ($a_{\rm ini}=4\times10^3\rm{R_\odot}$) with an inner blue line ($M_{\rm AGB,ini}=1.42\rm{M_\odot}$) that is  solid, dashed and dotted ($M_{\rm WD}=0.7$, 1.0 and $1.25\rm{M_\odot}$). These show the rate of change in orbital separation to increases more substantially for a more massive WD, which is clear by the steeper slope for more a massive WD. This is because in the long run, our more massive WDs lose more mass, even though  they retain more mass at the end of each eruption. But since each eruption still leads to a net mass loss, and they experience much more eruptions due to their lower triggering mass, they lose more mass, which leads to a more rapid widening of the orbit.

\subsection{What is the effect of mass change on the angular momentum of the system?}

To analyze the AML sinks, we examine the major sources, which are due to: mass transfer or loss ($\dot{J}_{\rm MT}$), magnetic braking ($\dot{J}_{\rm MB}$), gravitational radiation ($\dot{J}_{\rm GR}$), and as the result of drag being inflicted on the WD that is embedded in the wind that has escaped ($\dot{J}_{\rm D}$). While the latter three always carry a negative sign and lead to orbital reduction (e.g., \citealt{2011MNRAS.415.1907M,2015ApJS..220...15P,2020NatAs...4..886H,2021MNRAS.501..201H,2024MNRAS.527.4806V}), the former can carry either a positive or negative sign. For cases where the mass transfer efficiency ($\zeta_{BHL}$) is low --- as in the BHL mechanism, and all our models presented in this work --- this component is positive, and always has a larger absolute value than the sum of the other components. To demonstrate this, we show the absolute value of the ratio of the mass transfer component ($\dot{J}_{\rm MT}$) to the sum of the three negative terms ($\dot{J}_{\rm NEG}$) 
in Figure \ref{fig:Fig8}, where for the range of our models, this ratio spans from about $\times70$ until about $\times10$, while higher accretion rates generally lead to a higher ratio --- because a higher accretion rate inadvertently means a higher wind rate, i.e., mass loss rate, as well. 
The Figure also shows that less massive WDs and larger separations generally have a higher ratio, while the role of the AGB mass is less pronounced. 
This is because, less massive WDs have a smaller accretion radius, and wider separations lead to less dense wind at the WD location --- hence they both lead to a higher ratio. The AGB mass, while showing a subtle influence here, has an indirect effect by controlling the total amount of mass lost and the wind rate --- which are both higher for more massive AGBs. All in all, this figure demonstrates that the key influencer here is the orbital separation.

\subsection{The secularly evolving WD mass}

Our results show the WDs mass to secularly decrease  for all our models, while the more massive WDs, which experienced a higher accretion rate, and therefore a better retention efficiency \textit{per cycle}, over the lifetime of the given AGB, ejects the most mass. This may be seen in Figure \ref{fig:Fig9} where we show, for each model, the total amount of accreted mass and ejected mass over the entire simulation ($m_{\rm acc,tot}$ and $m_{\rm ej,tot}$) i.e., over the lifetime of the AGB. The total evolutionary trend is the opposite of the cyclic trend: per cycle, a less massive WD accretes more mass than a more massive one, while on an evolutionary perspective, it is the more massive WD that accretes more mass. This is because for given AGB masses and separations, the more massive WD has a higher accretion rate, thus manages to capture more mass than the less massive WDs. The total ejected mass also shows an opposite trend, and this is because the accretion rate, while varies between models due to varying accretion efficiency, is still below the mass-gaining threshold, thus for each nova cycle the WD will eject more mass than it has accreted. The recent studies on SySs such as U Sco \cite[]{2025ApJ...991..110S} and T CrB \cite[]{2025ApJ...991..111S} imply that both WDs eject more mass during nova eruptions than they have accreted since the last eruption. This means the WDs' masses are secularly decreasing—similar to our results—indicating that the BHL accretion mechanism is insufficient to grow the WD’s mass, rendering it an implausible type Ia supernova progenitor.

\section{Conclusions}\label{sec:Conclusions}

In this work, we have investigated the evolution of SySs that produce nova eruptions via various combinations of three basic systemic initial parameters: the WD mass, the donor AGB mass, and the orbital separation. 
We focused this work on the parameter regime for which the dominant mass transfer mechanism is the BHL mechanism as calculated by \cite{2025ApJ...980..224V}. We included one hypothetical case with a substantial smaller separation, which is below the threshold for BHL accretion \cite[]{2025ApJ...980..224V} in order to better understand the trends emerging from this investigation. 
Since our aim was to isolate the affect of each parameter on the systems evolution, we performed the analysis in three parts, each part using constant initial input for two of the three parameters, while the third parameter was varied. Our results revealed a number of trends, as we describe below. 

\textit{We find that, for a given wind rate, the accretion rate is higher for more massive WDs and for smaller orbits} --- in the former case due to their stronger gravity capturing more of the wind, and in the latter case due to enhanced mass tranfer at smaller separations. In contrast, \textit{the AGB mass does not directly affect the accretion rate, although it has an indirect effect via the wind rate}, which varies with AGB mass and changes drastically throughout its lifetime due to thermal pulses.

We note that the effect of the WD mass on the evolution of the donor in wind accretion dominated SySs operates in an entirely different way than it does in RLOF dominated CVs. In both CVs, and SySs an increased WD mass enhances the accretion rate,  however, in CVs, since the mass transfer in typically via RLOF, a massive WD erodes the donor at a faster rate since its gravity pulls the mass from the donor at a higher rate, whereas in wind accretion in SySt, the WD has no influence on the lifetime of the AGB donor. This is because, while RLOF mass transfer is driven by the WD’s gravity pulling material from the donor through the inner Lagrangian point --- a higher accretion rate thus leads to faster erosion of the donor, resulting in a shorter evolutionary timescale \citep{2021MNRAS.505.3260H}, whereas in wind accretion via BHL the evolutionary timescale is determined solely by the donor, regardless of the WD accretor, which simply captures some of the mass that the donor has already shed through its stellar wind.

Additionally, we find that more massive WDs experience \textit{a higher total mass loss despite having higher accretion efficiencies}. This counterintuitive outcome arises because their relatively low accretion rates --- though still comparatively higher than those of less massive WDs --- permit better mass retention after each cycle compared to lower mass WDs, where more extensive mixing leads to more mass being expelled during each nova eruption. However, this comparatively higher accretion rate in more massive WDs results in \textit{more frequent eruptions}, increasing cumulative mass loss compared to lower mass WDs, culminating in a higher total evolutionary mass loss.

Following this, We also find that our WD models that are paired with AGBs that have \textit{longer evolutionary times, experience more total mass loss} as well, indicating that the wind rate plays an important role in determining the retention efficiency ($\eta$) by determining  the accretion rate.

\textit{All the models} employed in this simulation showed a \textit{significant increase in the binary separation} over time --- that is, the orbit \textit{always} widens. This widening occurs because of the low mass transfer efficiency ($\zeta_{\rm BHL}$) that is yielded by the BHL mass transfer mechanism, as well as the almost always negative mass retention efficiency ($\eta$),  indicating a substantial mass loss from the system --- not only via stellar wind but also through nova eruptions. Although drag forces typically cause orbital contraction in binaries \cite{2024MNRAS.527.4806V}, here the angular momentum loss from mass transfer outweighs drag effects, making drag negligible and allowing the orbit to expand, while even though the orbit expands and causes the mass transfer efficiency to decrease, within the lifetime of the donor in the AGB phase, the average accretion rate remains in the range $10^{-9}-10^{-10}~ \rm{M_\odot~{yr}^{-1}}$, allowing periodic nova eruptions.

The relative change in orbital separation is \textit{higher} for: (1) donors that lose more mass --- i.e., initially \textit{more massive AGBs}; (2) initially \textit{wider} orbits; and (3) \textit{less massive WDs}. These conditions lead to more mass lost from the system, resulting in more substantial widening of the orbit. 
In line with this, the \textit{overall change in orbital separation is higher for more massive WDs}.

The final fate of these systems would be a double WD system, after the AGB completes its transformation. We do not see cases where the simulation suggests a possible shift to mass transfer regime of wind-RLOF because the orbit widens consistently. However, the reverse transition --- from the wind-RLOF mechanism to the BHL mechanism --- remains viable and we reserve the detailed investigation of such cases for future research.
 
\section*{Acknowledgments}
We thank the anonymous referee for the very helpful comments. IBV acknowledges support from the AGASS Center at Ariel University. We also acknowledge support from the Ariel University Research and Development Authority. Computational resources were provided by the Ariel HPC Centre at Ariel University, which we gratefully acknowledge. YH acknowledges the support of the Azrieli College of
Engineering - Jerusalem Research Fund. 

\section*{Data availability}
The data co-related with this work will be shared on reasonable request to the corresponding author.

\appendix

\section{Varying AGB mass---Additional Figures}\label{A1}
\begin{figure*}\renewcommand\thefigure{A1}
    \includegraphics[trim={1.0cm 1.0cm 1.0cm 1.0cm},clip,width=1\columnwidth]{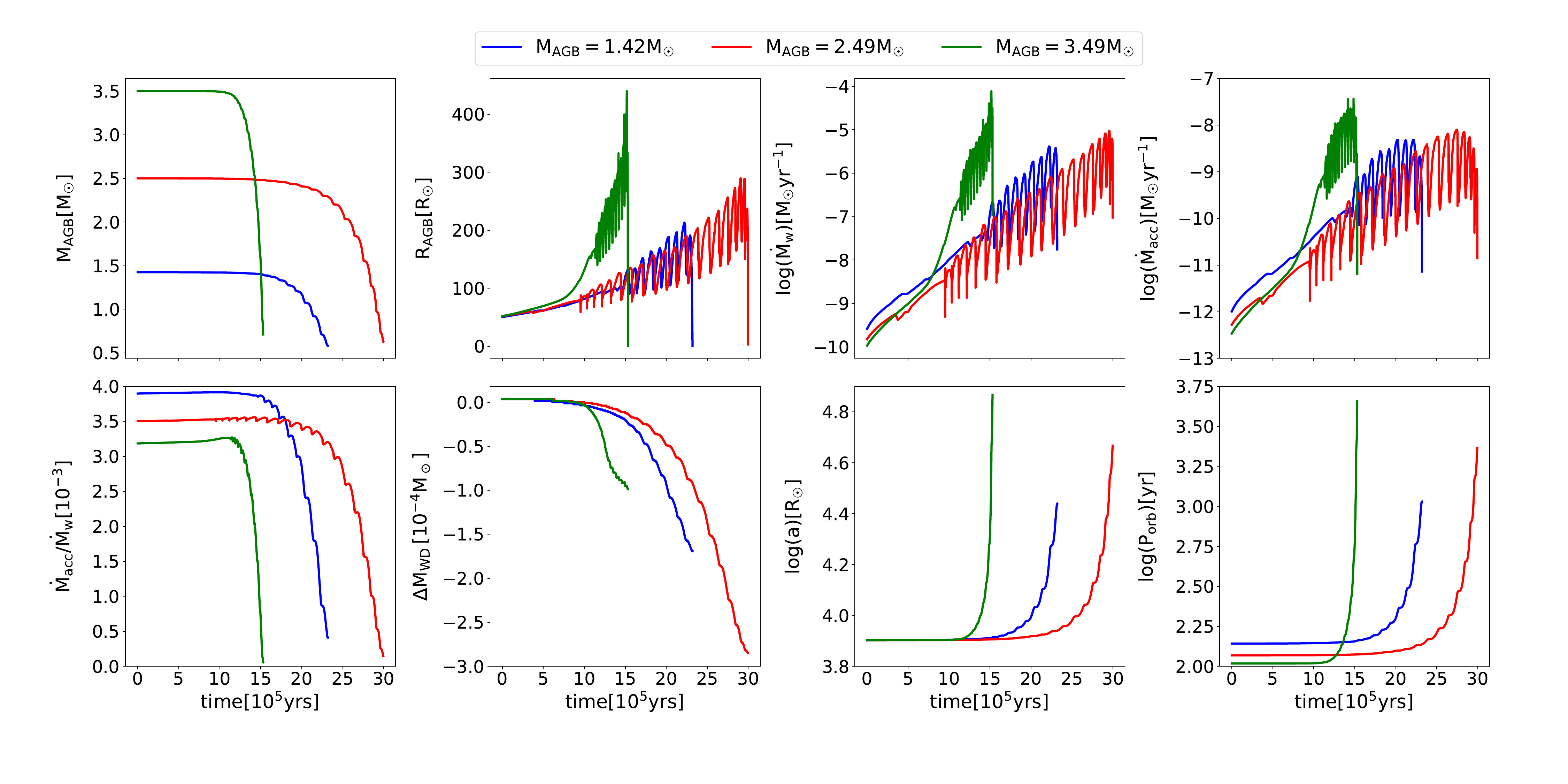}
    \caption{The figure follows the same format as described in Figure \ref{Fig1}, for a 1.25$M_\odot$ WD. }
    \label{fig:A1}
\end{figure*}

\begin{figure*}\renewcommand\thefigure{A2}
    \includegraphics[trim={1.0cm 1.0cm 1.0cm 1.0cm},clip,width=1\columnwidth]{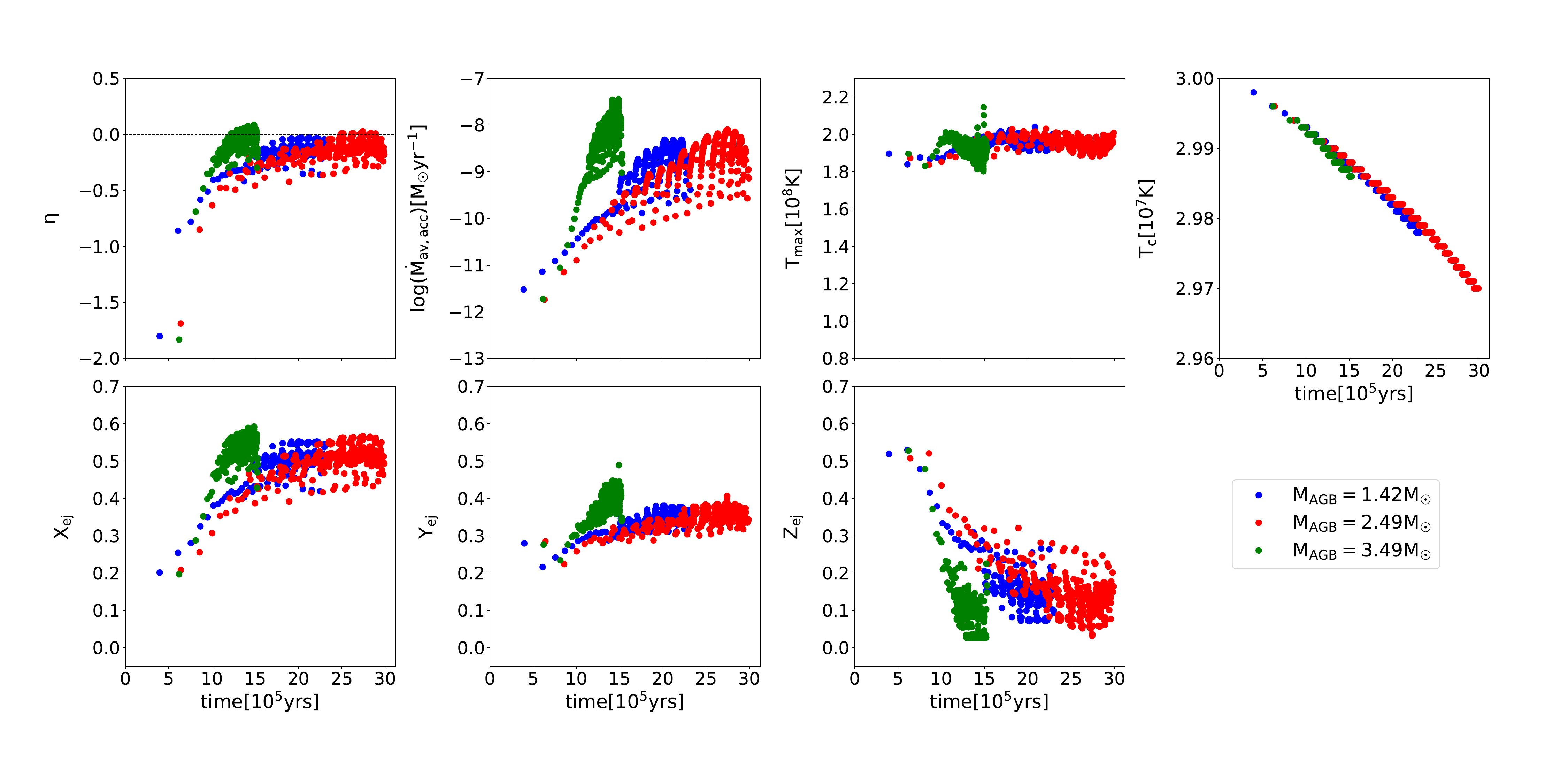}
    \caption{The figure follows the same format as described in Figure \ref{Fig2}, for a 1.25$M_\odot$ WD. }
    \label{fig:A2}
\end{figure*}

\begin{figure*}\renewcommand\thefigure{A3}
    \includegraphics[trim={1.0cm 1.0cm 1.0cm 1.0cm},clip,width=1\columnwidth]{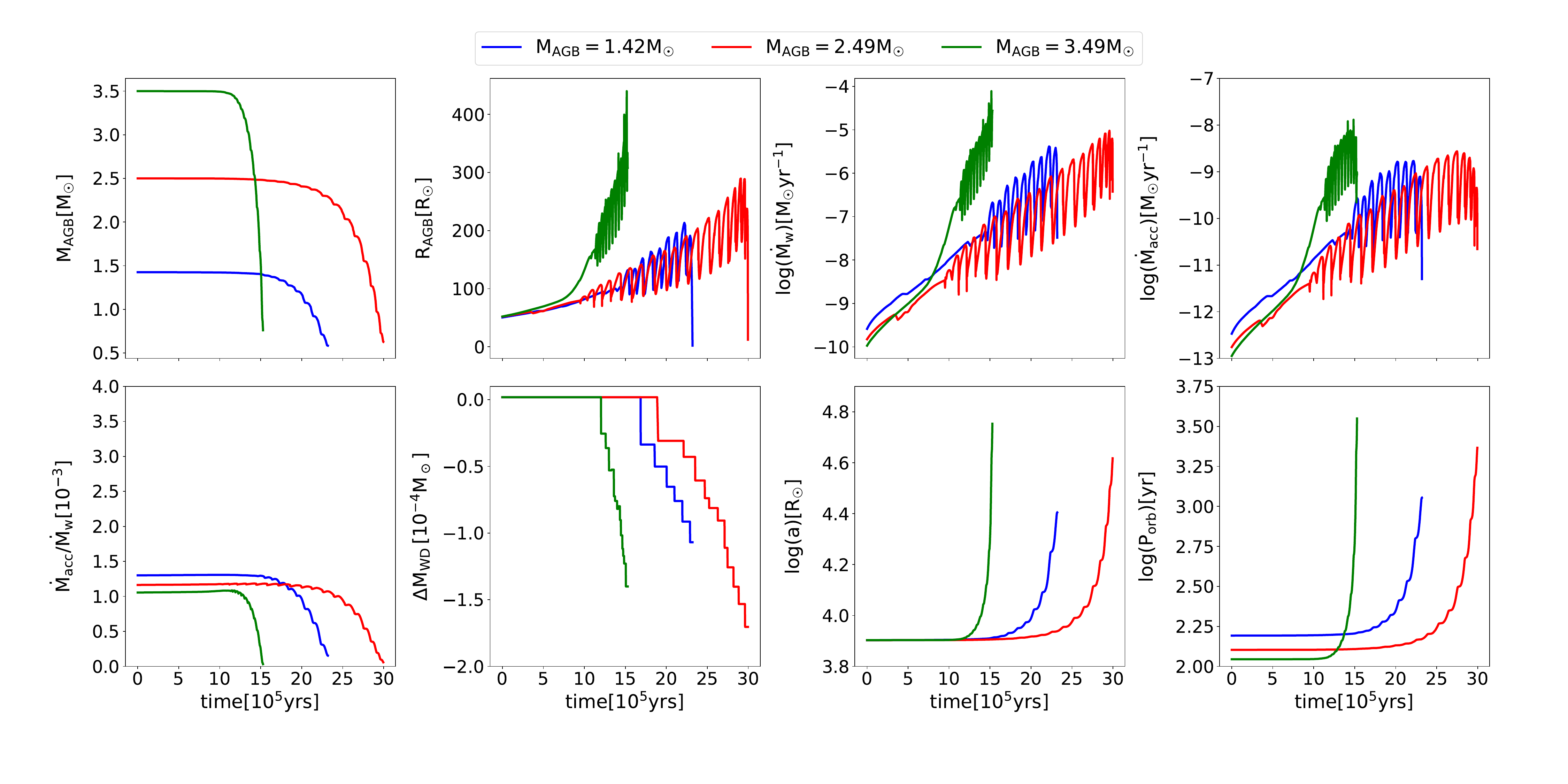}
    \caption{The figure follows the same format as described in Figure \ref{Fig1}, for a 0.7$M_\odot$ WD. }
    \label{fig:A3}
\end{figure*}

\begin{figure*}\renewcommand\thefigure{A4}
    \includegraphics[trim={1.0cm 1.0cm 1.0cm 1.0cm},clip,width=1\columnwidth]{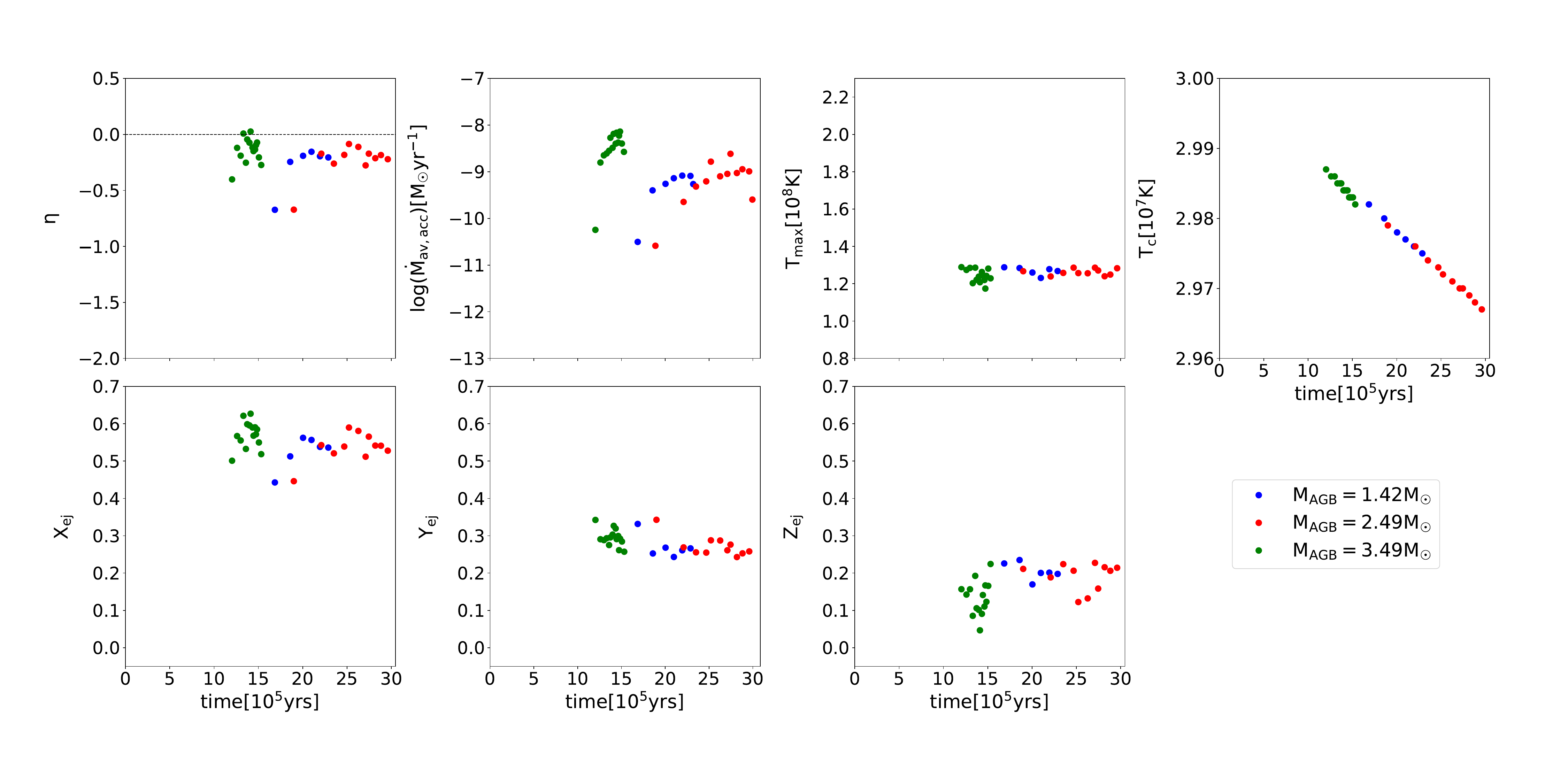}
    \caption{The figure follows the same format as described in Figure \ref{Fig2}, for a 0.7$M_\odot$ WD. }
    \label{fig:A4}
\end{figure*}

Figures \ref{fig:A1}, \ref{fig:A3} and \ref{fig:A2}, \ref{fig:A4} show the evolution of parameters as described in Figures \ref{Fig1} and \ref{Fig2} respectively, for WDs with masses of 1.25$M_\odot$ and 0.7$M_\odot$. The general behavior of all the parameters remains the same, though a better mass transfer efficiency is obtained for more massive WDs. Similarly, an increase in $T_{\rm max}$ is obtained for more massive WDs as well, due to their smaller, more compact size and stronger gravitational binding, which raises the maximum temperature during nova eruptions.

\section{Varying Separation---Additional Figures}\label{B1}

Figures \ref{fig:B1} and \ref{fig:B2} show the evolution of parameters for two different separations (6000$\rm{R_{\odot}}$ and 8000$\rm{R_{\odot}}$) in a system with an initial AGB mass 2.49$\rm{M_{\odot}}$ and WD mass of 1.0$\rm{M_{\odot}}$, following the same description as in Figures \ref{Fig3} and \ref{Fig4}, respectively. The initial separation of 6000$\rm{R_{\odot}}$ is obtained from the calculations by \cite{2025ApJ...980..224V}, ensuring that the system falls within the BHL accretion regime. The general behavior remains consistent with what is detailed in Section \ref{32}. The only change is in the mass transfer efficiency, which alters the accretion rate and the evolution of orbital separation and orbital period, although the system still follows the general pattern.

\begin{figure*}\renewcommand\thefigure{B1}
    \includegraphics[trim={1.0cm 1.0cm 1.0cm 1.0cm},clip,width=1\columnwidth]{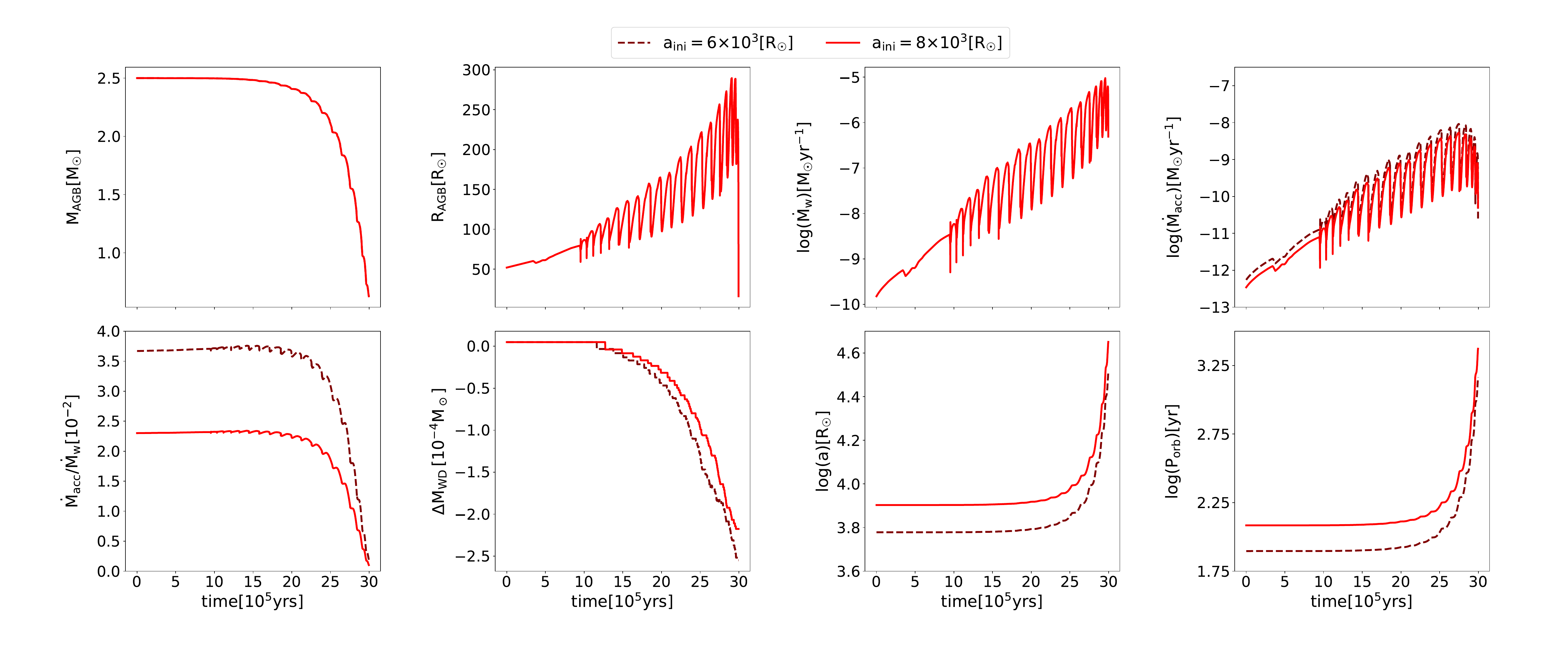}   
    \caption{The figure follows the same format as described in Figure \ref{Fig3}, for a 2.49$M_\odot$ AGB with an initial separation of a=8000$\rm{R_\odot}$. }
    \label{fig:B1}
\end{figure*}

\begin{figure*}\renewcommand\thefigure{B2}
    \includegraphics[trim={1.0cm 1.0cm 1.0cm 1.0cm},clip,width=1\columnwidth]{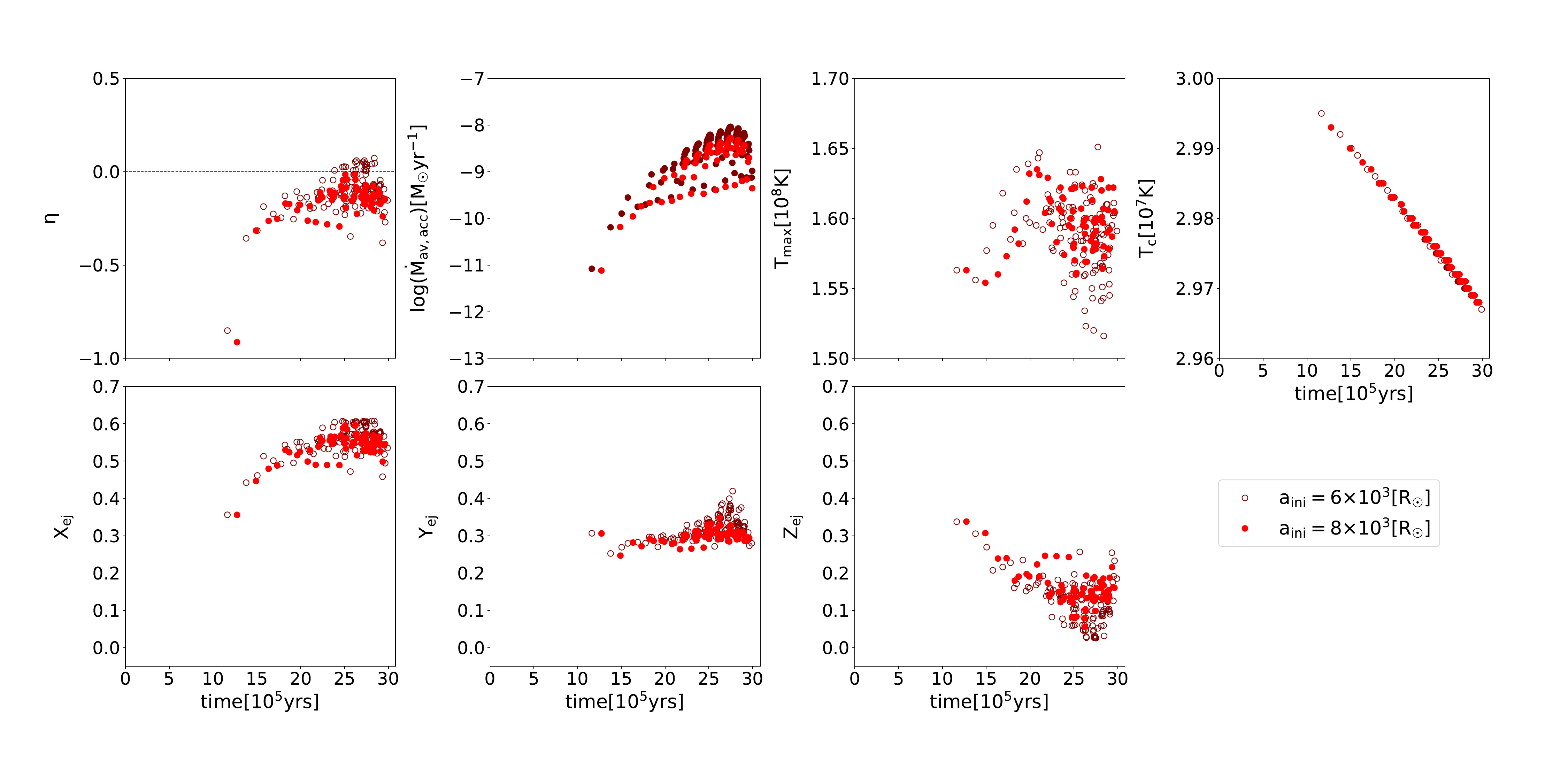}    
    \caption{The figure follows the same format as described in Figure \ref{Fig4}, for a 2.49$M_\odot$ AGB with an initial separation of a=8000$\rm{R_\odot}$. }
    \label{fig:B2}
\end{figure*}

\section{Varying WD mass---Additional Figures}\label{C1}

Figures \ref{fig:C1}, \ref{fig:C3} and \ref{fig:C2}, \ref{fig:C4} show the evolution of parameters for initial AGB masses of 2.49$\rm{M_{\odot}}$ and 3.49$\rm{M_{\odot}}$ on different WD mass with initial separation of 8000$\rm{R_{\odot}}$, following the same descriptions as in Figures \ref{Fig5} and \ref{Fig6}, respectively. The evolution of these parameters follows the trend explained in Section \ref{33}.

\begin{figure*}\renewcommand\thefigure{C1}
    \includegraphics[trim={1.0cm 1.0cm 1.0cm 1.0cm},clip,width=1\columnwidth]{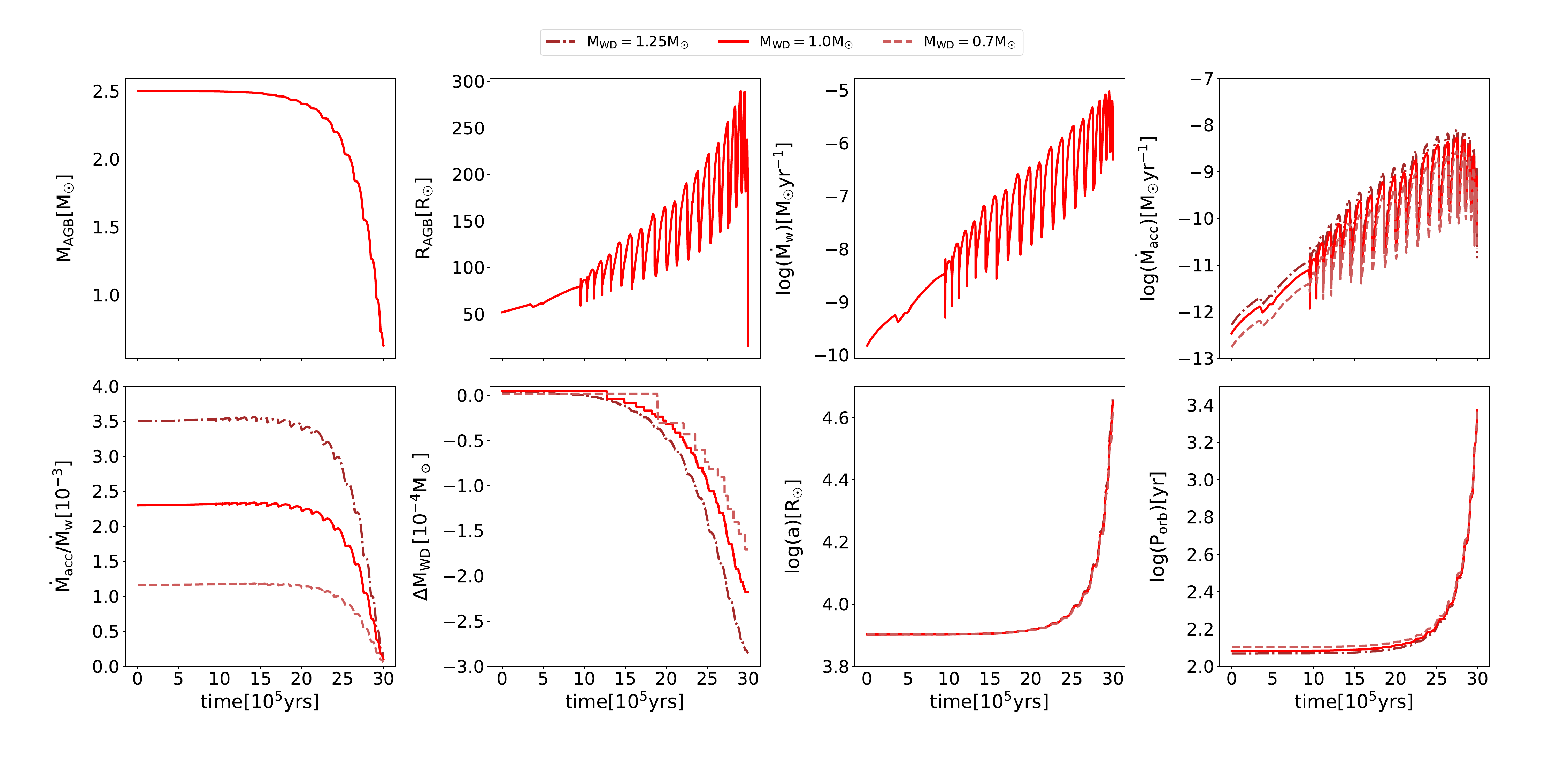}
    \caption{The figure follows the same format as described in Figure \ref{Fig5}, for a 2.49$M_\odot$ AGB with an initial separation of a=8000$\rm{R_\odot}$. }
    \label{fig:C1}
\end{figure*}

\begin{figure*}\renewcommand\thefigure{C2}
    \includegraphics[trim={1.0cm 1.0cm 1.0cm 1.0cm},clip,width=1\columnwidth]{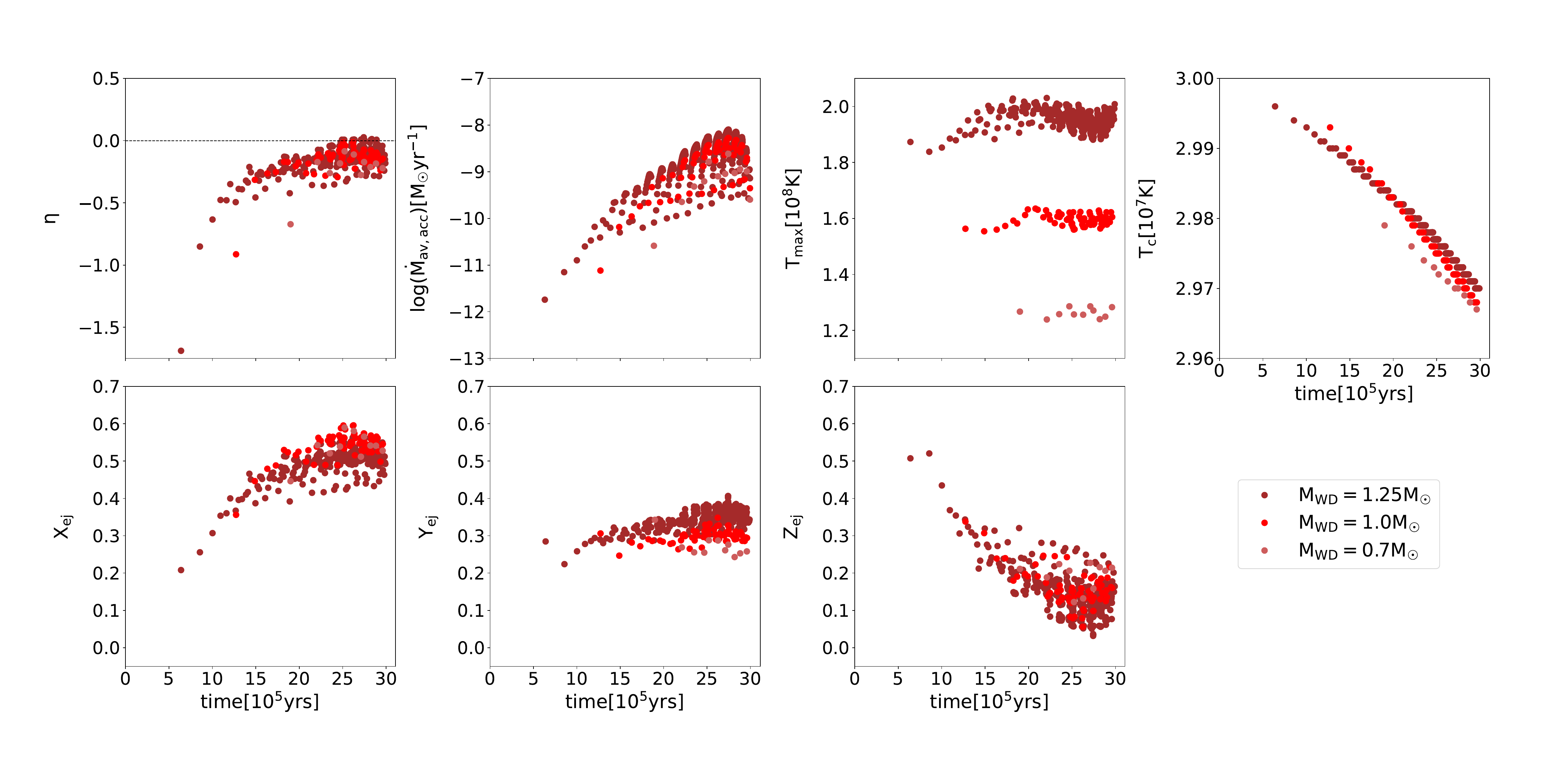}
    \caption{The figure follows the same format as described in Figure \ref{Fig6}, for a 2.49$M_\odot$ AGB with an initial separation of a=8000$\rm{R_\odot}$. }
    \label{fig:C2}
\end{figure*}

\begin{figure*}\renewcommand\thefigure{C3}
    \includegraphics[trim={1.0cm 1.0cm 1.0cm 1.0cm},clip,width=1\columnwidth]{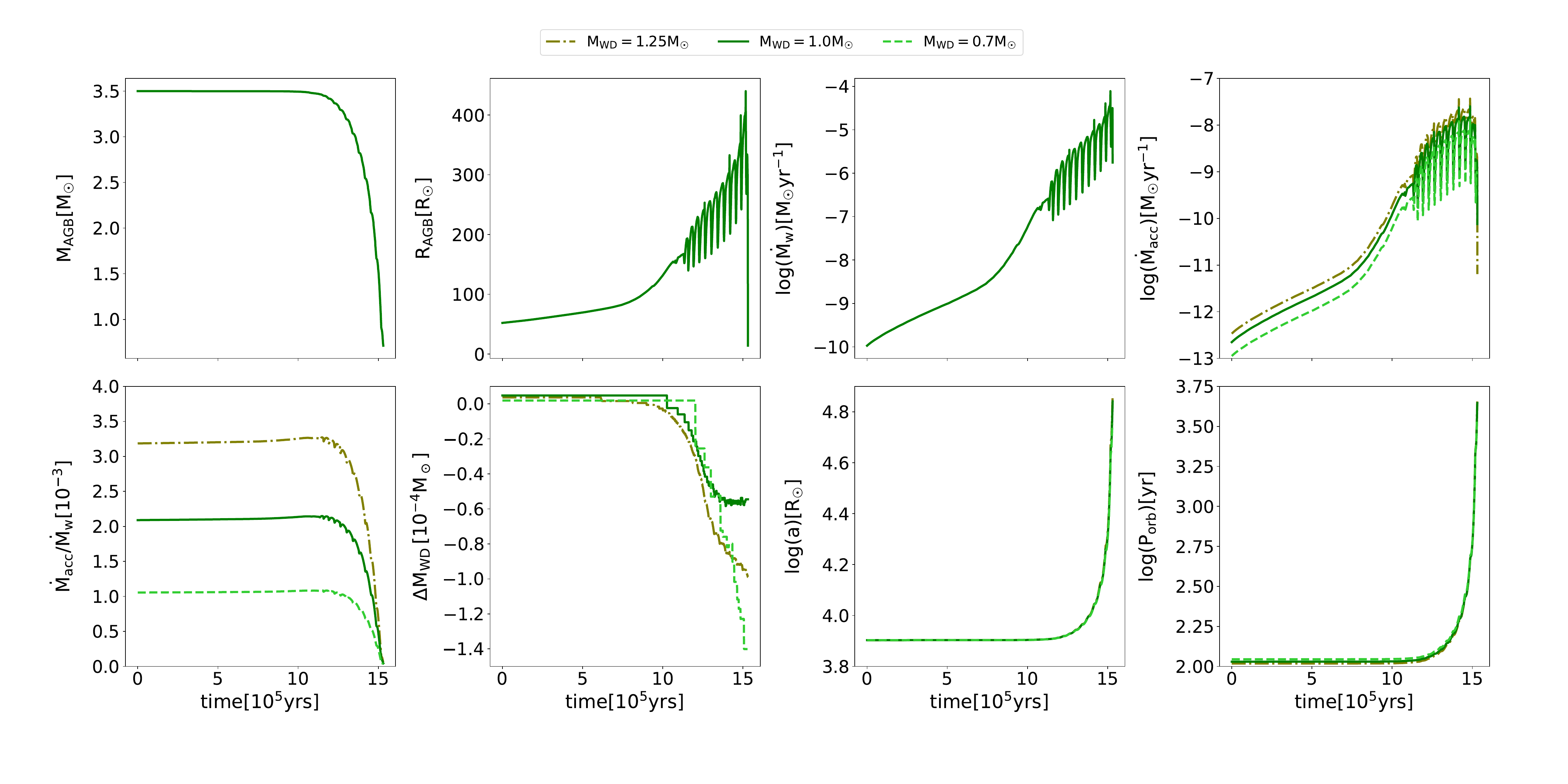}
    \caption{The figure follows the same format as described in Figure \ref{Fig5}, for a 3.49$M_\odot$ AGB with an initial separation of a=8000$\rm{R_\odot}$. }
    \label{fig:C3}
\end{figure*}

\begin{figure*}\renewcommand\thefigure{C4}
    \includegraphics[trim={1.0cm 1.0cm 1.0cm 1.0cm},clip,width=1\columnwidth]{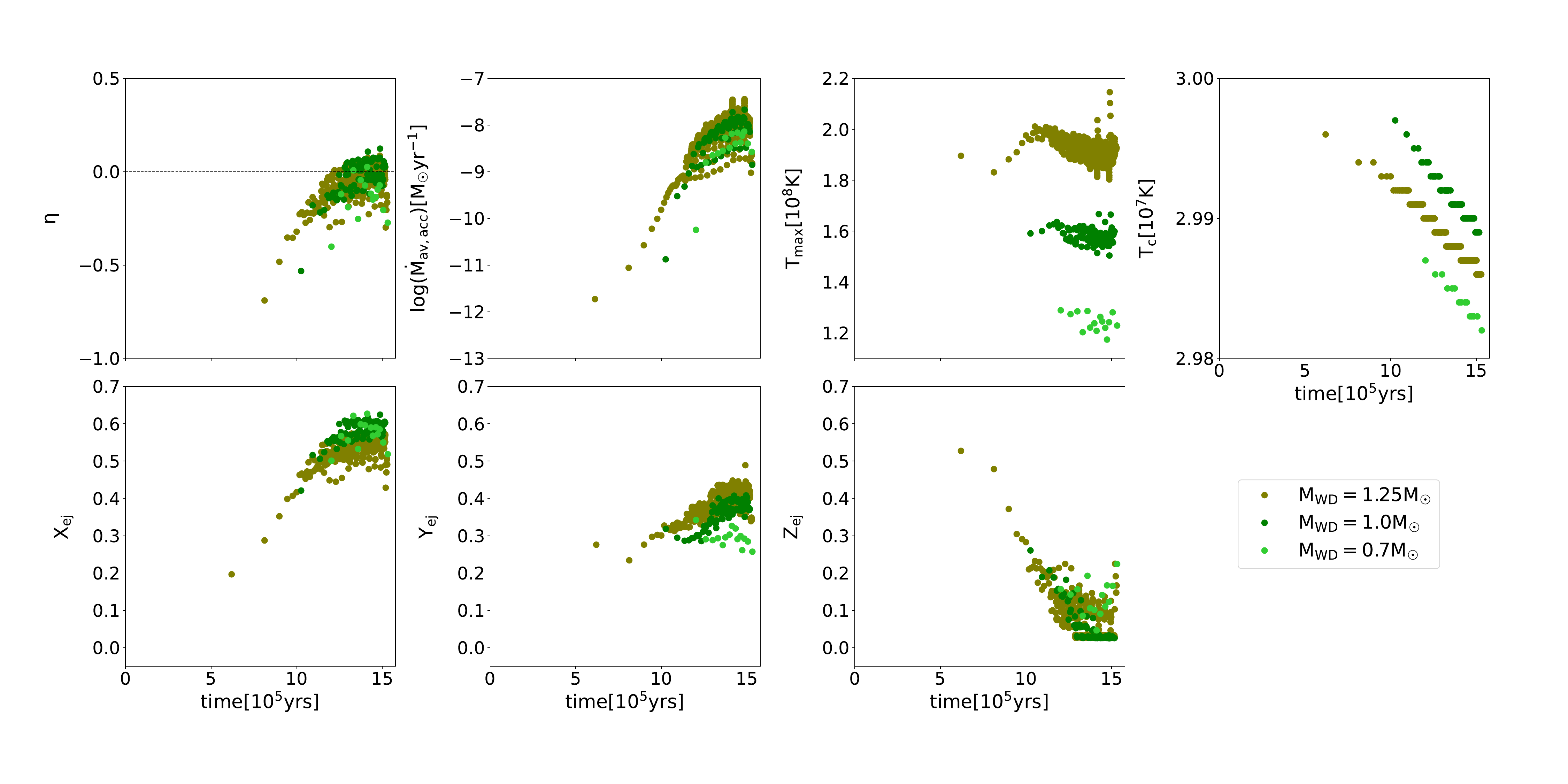}
    \caption{The figure follows the same format as described in Figure \ref{Fig6}, for a 3.49$M_\odot$ AGB with an initial separation of a=8000$\rm{R_\odot}$. }
    \label{fig:C4}
\end{figure*}

\bibliography{ref}{}
\bibliographystyle{aasjournalv7}

\end{document}